\documentclass[11pt]{article}
\usepackage{amssymb,amsmath,amsthm,epsfig,hyperref}

\newcommand{\treee}[5]{
\begin{picture}(80,60)(0,0)
\put(10,9){\makebox(0,15){#2}} \put(10,23){\line(1,1){10}}
\put(70,43){\line(-1,-1){10}} \put(70,23){\line(-1,1){10}}
\put(10,43){\line(1,-1){10}} \put(20,33){\line(1,0){40}}
\put(40,33){\line(0,-1){10}} \put(40,9){\makebox(0,15){#3}}
\put(70,9){\makebox(0,15){#4}} \put(10,41){\makebox(0,15){#1}}
\put(70,41){\makebox(0,15){#5}}
\end{picture}}

\def\Prob#1{{\mathbf{Pr}\left({#1}\right)}}

\def\ProbCond#1#2{{\mathbf{Pr}\left({#1} \mid {#2} \right)}}

\newtheorem{theorem}{Theorem}

\newtheorem{pro}[theorem]{Proposition}
\newtheorem{corollary}[theorem]{Corollary}
\newtheorem{observation}[theorem]{Observation}
\newtheorem{lemma}[theorem]{Lemma}
\newtheorem{lem}[theorem]{Lemma}

\newtheorem{obs}[theorem]{Observation}
\newtheorem{dfn}[theorem]{Definition}
\newtheorem{definition}[theorem]{Definition}
\newtheorem{remark}[theorem]{Remark}

\newcommand{\eps}{\epsilon}
\newcommand{\Dvector}{\overrightarrow{D}}

\def\ra{\rightarrow}
\def\BPRO{\begin{pro}}
\def\EPRO{\end{pro}}
\def\BOBS{\begin{obs}}
\def\EOBS{\end{obs}}
\def\BPRF{\begin{proof}}
\def\EPRF{\end{proof}}
\def\BLEM{\begin{lem}}
\def\ELEM{\end{lem}}
\def\BDEF{\begin{dfn}\rm}
\def\EDEF{\end{dfn}}
\def\BCOR{\begin{corollary}}
\def\ECOR{\end{corollary}}

\def\A{{\sf{A}}}
\def\G{{\sf{G}}}
\def\T{{\sf{T}}}
\def\C{{\sf{C}}}
\def\intg{\mathbb{Z}}
\def\<{\langle}
\def\>{\rangle}

\newcommand{\vecD}{\overrightarrow{D}}
\newcommand{\Dvec}{\overrightarrow{D}}

\renewcommand{\vec}[1]{{\overrightarrow{#1}}}
\renewcommand{\vector}[1]{\overrightarrow{#1}}

\newcommand{\xhat}{\widehat{x}}
\newcommand{\yhat}{\widehat{y}}
\newcommand{\zhat}{\widehat{z}}
\newcommand{\what}{\widehat{w}}

\def\real{\mathbb{R}}

\title{Phylogeny of Mixture Models:
Robustness of Maximum Likelihood and Non-identifiable Distributions}

\author{Daniel \v{S}tefankovi\v{c}\thanks{
Department of Computer Science, University of Rochester,
Rochester, NY 14627, and Comenius University, Bratislava.  Email: stefanko@cs.rochester.edu}
\and Eric Vigoda\thanks{College of Computing, Georgia
Institute of Technology, Atlanta GA 30332.  Email: vigoda@cc.gatech.edu.
Research supported in part by NSF grant CCF-0455666} }

\date{September 25, 2006}

\begin{document}

\maketitle

\begin{abstract}

We address phylogenetic reconstruction when the data is generated from a mixture distribution.
Such topics
have gained considerable attention in the biological community with the
clear evidence of heterogeneity of mutation rates.
 In our work we consider data coming from a mixture of trees which share a
 common topology, but differ in their edge weights (i.e., branch lengths).
 We first show the pitfalls of popular methods, including maximum
 likelihood and Markov chain Monte Carlo algorithms.
 We then determine in which evolutionary
 models, reconstructing the tree topology, under a mixture
 distribution, is (im)possible.
We prove that every model whose transition matrices
can be parameterized by an open set of multi-linear polynomials, either
has non-identifiable mixture distributions, in which case reconstruction
is impossible in general, or there exist linear tests which identify the topology.  This
duality theorem, relies on our notion
of linear tests and uses ideas from convex programming duality.  Linear tests are
closely related to linear invariants, which were first introduced by Lake,
and are natural from an algebraic geometry perspective.

\end{abstract}

\section{Introduction}

A major obstacle to phylogenetic inference is the heterogeneity of
genomic data.  For example, mutation rates vary widely between genes,
resulting in different branch lengths in the phylogenetic tree for
each gene.  In many cases, even the topology of the tree differs between genes.
Within a single long gene, we are also likely to see
variations in the mutation rate, see \cite{Hellmann2005} for a current
study on regional mutation rate variation.

Our focus is on phylogenetic inference based on single nucleotide
substitutions.  In this paper we study the effect of mutation rate
variation on phylogenetic inference.
The exact mechanisms of single nucleotide substitutions are still
being studied,
 hence  the causes of variations in the rate of these mutations are unresolved, see
 \cite{Hellmann2003,Webster2004,Meunier2004}.
In this paper we study phylogenetic inference in the presence of heterogeneous data.

For homogenous data, i.e., data generated from a single
phylogenetic tree, there is considerable work on
consistency of various methods, such as likelihood \cite{Chang:cons} and distance methods,
and inconsistency of other methods, such as
parsimony \cite{F1978}.  Consistency means that the methods converge
to the correct tree with sufficiently large amounts of data.
We refer the interested reader to Felsenstein \cite{F:book}
for an introduction to these phylogenetic approaches.

There are several works showing the pitfalls of
popular phylogenetic methods when data is generated
from a mixture of trees, as opposed to a single tree.
We review these works in detail shortly.
The effect of mixture distributions has been of marked interest
recently in the biological community, for instance, see
the recent publications of Kolczkowski and Thornton \cite{KT:nature},
and Mossel and Vigoda \cite{MV}.

In our setting the data is generated from a mixture of trees which have a common
tree topology, but can vary arbitrarily in their mutation rates.  We address whether it
is possible to infer the tree topology.
We introduce the notion of a {\em linear test}.  For any mutational
model whose transition probabilities can be parameterized by an open set
(see the following subsection for a precise definition),
we prove that the topology can be reconstructed by linear tests, or it is
impossible in general due to a non-identifiable mixture distribution.
For several of the popular mutational models
we determine which of the two scenarios (reconstruction or non-identifiability) hold.

The notion of a linear test is closely related
to the notion of linear invariants.  In fact, Lake's invariants are a linear
test.  There
are simple examples where linear tests exist and linear invariants do not
(in these examples,  the mutation rates are restricted to
some range).
However, for the popular mutation models, such as Jukes-Cantor and
Kimura's 2 parameter model (both of which are
closed under multiplication) we have no such examples.
For the Jukes-Cantor and Kimura's 2 parameter model, we prove the linear tests are essentially
unique (up to certain symmetries).
 In contrast to
the study of invariants,  which is natural from an algebraic geometry
perspective, our work is based on convex programming duality.

We present the background material before formally stating our new results.
We then give a detailed comparison of our results with related previous work.

An announcement of the main results of this paper, along with some
applications of the technical tools presented here,  are presented in \cite{SV:sysbio}.

\subsection{Background}

A phylogenetic tree is an unrooted tree $T$ on $n$ leaves (called taxa, corresponding to
$n$ species) where
internal vertices have degree three.
Let $E(T)$ denote the edges of $T$ and $V(T)$ denote the vertices.
The mutations along edges of $T$ occur according
to a continuous-time Markov chain.  Let $\Omega$ denote the states of the model.
The case $|\Omega|=4$ is biologically important, whereas
$|\Omega|=2$ is mathematically convenient.

The model is defined by a phylogenetic tree $T$ and a distribution $\pi$ on $\Omega$.
Every edge $e$
has an associated $|\Omega|\times |\Omega|$ rate matrix $R_e$, which is reversible with respect
to $\pi$, and a time $t_e$.   Note, since $R_e$ is reversible with respect to $\pi$, then
$\pi$ is the stationary vector for $R_e$ (i.e., $\pi R_e = 0$).
The rate matrix defines a continuous time Markov chain.  Then, $R_e$ and $t_e$ define
a transition matrix $P_e=\exp(t_eR_e)$.
The matrix is a stochastic matrix of size $|\Omega|\times|\Omega|$, and thus defines a discrete-time
Markov chain, which is time-reversible, with stationary distribution $\pi$ (i.e., $\pi P_e = \pi$).

Given $\vector{P}=(P_{e})_{e\in E(T)}$
we then define the following distribution on labellings of the vertices of $T$.
We first orient the edges of $T$ away from an arbitrarily chosen root
$r$ of the tree.  (We can choose the root arbitrarily since each $P_e$ is reversible with
respect to $\pi$.)   Then,
the probability of a labeling
$\ell:V(T)\ra\Omega$ is
\begin{equation}
\label{eq:defp}
\mu'_{T,\vec{P}}(\ell)=\pi(\ell(r))\prod_{\overrightarrow{uv}\in E(T)}
\vector{P}_{uv}(\ell(u),\ell(v)).
\end{equation}

Let $\mu_{T,\vec{P}}$ be the marginal distribution of $\mu'_{T,\vec{P}}$ on
the labelings of the leaves of $T$ ($\mu_{T,\vec{P}}$ is a distribution
on $\Omega^n$ where $n$ is the number of leaves of $T$). The goal of
phylogeny reconstruction is to reconstruct $T$ (and possibly $\vec{P}$)
from $\mu_{T,\vec{P}}$ (more precisely, from independent samples from $\mu_{T,\vec{P}}$).

The simplest four-state model has a single parameter $\alpha$ for the
off-diagonal entries of the rate matrix.  This model is known as the Jukes-Cantor
model, which we denote as JC.
Allowing 2 parameters in the rate matrix is Kimura's 2 parameter model
which we denote as K2, see Section \ref{sec:K2} for a formal definition.
 The K2 model accounts for the higher mutation rate of
transitions (mutations  within
purines or pyrimidines) compared to transversions (mutations between a
purine and a pyrimidine).
Kimura's 3 parameter model, which we refer to as K3
accounts for the number of hydrogen bonds altered by the mutation.
See Section \ref{sec:K3} for a formal definition of the K3 model.
For $|\Omega|=2$, the model is binary and the rate matrix has a single parameter $\alpha$.
This model is known as the CFN (Cavender-Farris-Neyman) model.
For any examples in this paper involving
the CFN, JC, K2 or K3 models, we restrict the model to rate matrices where
all the entries are positive, and times $t_e$ which are positive and finite.

We will use $\mathcal{M}$ to denote the set of transition matrices obtainable by the model
under consideration, i.e.,
\[  \mathcal{M} = \{P_e=\exp(t_eR_e : t_e \mbox{ and } R_e \mbox{ are allowed in the
model }\}.
\]
The above setup allows additional restrictions in the model,
such as requiring $t_e>0$ which is commonly required.

In our framework, a model is specified by a set $\mathcal{M}$, and then each
edge is allowed any transition matrix $P_e\in\mathcal{M}$.
We refer to this framework as the {\em unrestricted framework},
since we are not imposing any dependencies on
the choice of transition matrices between edges.  This set-up is
convenient since it gives a natural algebraic framework for the
model as we will see in some later proofs.  A similar set-up was
required in the work of Allman and Rhodes \cite{AllmanRhodes}, also
to utilize the algebraic framework.

An alternative framework (which is typical in practical
works) requires a common rate
matrix for all edges, specifically $R=R_e$ for all $e$.
Note we can not impose such a restriction
in our unrestricted framework, since each edge is allowed any matrix in $\mathcal{M}$.
We will refer to this framework as the {\em common rate framework}.
Note, for the Jukes-Cantor and CFN models, the unrestricted and common rate frameworks
are identical, since there is only a single parameter for each edge in these models.
We will discuss how our results
apply to the common rate framework when relevant, but the default setting of our results
is the unrestricted model.

Returning to our setting of the unrestricted framework,
recall under the condition $t_e>0$
the set $\mathcal{M}$ is not a compact set
(and is parameterized by an open set as described shortly).
This will be important for our work
since our main result will only apply to models where
$\mathcal{M}$ is an open set.  Moreover we will require that $\mathcal{M}$
consists of multi-linear polynomials.  More precisely,
a polynomial $p\in\real[x_1,\dots,x_m]$ is {\em multi-linear} if for each
variable $x_i$ the degree of $p$ in $x_i$ is at most $1$.
Our general results will apply when the model is a set
of multi-linear polynomials which are parameterized by an
open set which we now define precisely.

\begin{definition}
We say that a set $\mathcal{M}$ of transition matrices is {\em
parameterized by an open set}
if there exists a finite set $\Omega$, a distribution $\pi$ over
$\Omega$, an integer $m$, an open set
$O\subseteq\real^m$, and multi-linear polynomials
$p_{ij}\in\real[x_1,\dots,x_m]$ such that
$$
{\cal M}=\{  (p_{ij})_{i,j=1}^{\Omega}\,|\, (x_1,\dots,x_m)\in O\},
$$
where $\mathcal{M}$ is a set of stochastic matrices which are reversible with
respect to $\pi$ (thus $\pi$ is their stationary distribution).
\end{definition}

Typically the polynomials $p_{ij}$ are defined by an appropriate change of
variables from the variables defining the rate matrices.  Some examples
of models that are paraemeterized by an open set are the general Markov model considered by
Allman and Rhodes \cite{AllmanRhodes}; Jukes-Cantor, Kimura's 2-parameter
and 3-parameter, and Tamura-Nei models.
For the Tamura-Nei model (which is a generalization
of Jukes-Cantor and Kimura's models) we show in \cite{SV:sysbio} how the model
can be re-parameterized in a straightforward manner
so that it consists of multi-linear polynomials, and
thus fits the parameterized by an open set condition (assuming the additional
restriction $t_e>0$).

 \subsection{Mixture Models}

In our setting, we will generate assignments from a mixture distribution.
We will have a single tree topology $T$, a collection of $k$ sets of transition matrices
$\vec{P_1},\vec{P_2},\dots,\vec{P_k}$ where $\vec{P_i}\in\mathcal{M}^{E(T)}$ and
a set of non-negative reals $q_1,q_2,\dots,q_k$ where $\sum_i q_i = 1$.
We then consider the mixture distribution:
\[
\mu = \sum_i q_i\mu_{T,\vec{P_i}}
\]
Thus, with probability $q_i$ we generate a sample according to
$\mu_{T,\vec{P_i}}$.  Note the tree topology is the same for all the distributions
in the mixture (thus there is a notion of a generating topology).
In several of our simple examples
we will set $k=2$ and $q_1=1/2$, thus we will be looking at a uniform
mixture of two trees.

\subsection{Maximum Likelihood and MCMC results}

We begin by showing a simple class of mixture distributions where
popular phylogenetic algorithms fail.  In particular, we consider
maximum likelihood methods, and Markov chain Monte Carlo (MCMC) algorithms for
sampling from the posterior distribution.

In the following, for a mixture distribution $\mu$,
we consider the likelihood of a tree $T$ as, the
maximum over assignments of transition matrices $\vec{P}$ to the edges of $T$,  of the
probability that the tree $(T,\vec{P})$ generated $\mu$.  Thus, we are considering the likelihood
of a pure (non-mixture) distribution having generated the mixture distribution $\mu$.
More formally, the maximum expected
log-likelihood of tree $T$ for distribution $\mu$ is defined by
\[
{\cal L}_{T}(\mu)  = \max_{\vec{P}\in{\mathcal{M}}^E} {\cal L}_{T,\vec{P}}(\mu),
\]
where
\[
 {\cal L}_{T,\vec{P}}(\mu) = \sum_{y\in\Omega^n} \mu(y)\ln(\mu_{T,\vec{P}}(y))
 \]

Recall for the CFN, JC, K2 and K3 models, $\mathcal{M}$ is restricted to
transition matrices obtainable from positive rate matrices $R_e$ and positive times $t_e$.

Chang \cite{Chang:incons} constructed a mixture example where likelihood (maximized over
the best single tree) was maximized on the wrong topology (i.e., different from the generating
topology).
In Chang's examples one tree had all edge weights sufficiently
small (corresponding to invariant sites).  We consider examples with
less variation within the mixture and fewer parameters required to be sufficiently small.
We consider (arguably more natural) examples of the same flavor as those
studied by Kolaczkowski and Thornton \cite{KT:nature},
who showed experimentally that in the JC model, likelihood appears to perform poorly
on these examples.

\begin{figure}[htb]
\begin{center}
\includegraphics[type=eps,ext=.eps,read=.eps,height=1in]{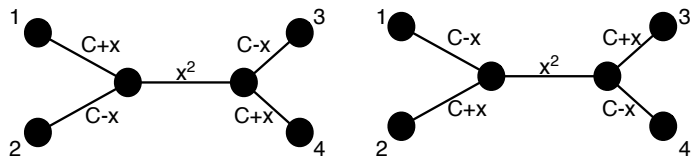}
\caption{In the binary CFN model, for all choices of $C$ and $x$ sufficiently small,
maximum likelihood is inconsistent on a mixture of the above trees.}\label{fig:CFNbadMLE}
\end{center}
\end{figure}

Figure \ref{fig:CFNbadMLE} shows the form of our examples where $C$ and $x$
are parameters of the example.  We consider a uniform mixture of the two trees in
the figure.  For each edge, the figure shows the mutation probability, i.e.,
it is the off-diagonal entry for the transition matrix.
We consider the CFN, JC, K2 and K3 models.

We prove that in this mixture model, maximum likelihood is not robust in the
following sense:  when likelihood is maximized over the best single tree, the
maximum likelihood topology is different from the generating topology.

In our example all of the off-diagonal entries of the transition matrices are identical.
Hence for each edge we specify a single parameter and thus we define
a set $\vector{P}$ of transition matrices for a 4-leaf tree by a 5-dimensional
vector where the $i$-th coordinate is the parameter for the edge incident leaf $i$,
and the last coordinate is the parameter for the internal edge.

Here is the statement of our result on the robustness of likelihood.
\begin{theorem}
\label{thm:MLEbad}
Let $C\in (0,1/|\Omega|)$.
Let  $\vector{P_1}=(C+x,C-x,C-x,C+x,x^2)$ and
$\vector{P_2}=(C-x,C+x,C+x,C-x,x^2)$.  Consider the following
mixture distribution on $T_3$:
\[ \mu_x=\left(\mu_{T_3,\vector{P_1}}+\mu_{T_3,\vector{P_2}}\right)/2.
\]
\begin{enumerate}
\item In the CFN model, for all $C\in(0,1/2)$,
there exists $x_0>0$ such that for all $x\in(0,x_0)$ the
maximum-likelihood tree for $\mu_x$ is $T_1$.
\label{thm:CFNbad}
\item In the JC, K2 and K3 models, for $C=1/8$,
there exists $x_0>0$ such that for all $x\in(0,x_0)$ the
maximum-likelihood tree for $\mu_x$ is $T_1$.
\label{thm:JCbad}
\end{enumerate}
Recall, likelihood is maximized over the best pure (i.e., non-mixture) distribution.
\end{theorem}

Note, for the above theorem, we are maximizing the likelihood over assignments of valid
transition matrices for the model.  For the above models, valid transition matrices are those
obtainable with finite and positive times $t_e$, and rate matrices $R_e$ where all the entries are positive.

A key observation for our proof approach of Theorem \ref{thm:MLEbad} is
that the two trees in the mixture example are the same in the limit $x\rightarrow 0$.
The $x=0$ case is used in the proof for the $x>0$ case.
We expect the above theorem holds for more a general class of examples
(such as arbitrary $x$, and any sufficiently small function on the internal edge),
but our proof approach requires $x$ sufficiently small.
Our proof approach builds upon the work of Mossel and Vigoda \cite{MV}.

Our results also extend to show, for the CFN and JC models,
MCMC methods using NNI transitions converge exponentially
slowly to the posterior distribution.  This result requires the 5-leaf version of mixture
example from Figure \ref{fig:CFNbadMLE}.  We state our MCMC result formally in
Theorem \ref{thm:MCMC-main} in Section \ref{sec:mcmc} after presenting the
background material.
Previously,
Mossel and Vigoda \cite{MV}
showed a mixture distribution where MCMC methods converge exponentially slowly to the
posterior distribution.  However, in their example, the tree topology varies between the two
trees in the mixture.

\subsection{Duality Theorem: Non-identifiablity or Linear Tests}

Based on the above results on the robustness of likelihood, we consider
whether there are any methods which are guaranteed to determine the common topology
for mixture examples.  We first found that in the CFN model there is a simple
mixture example of size $2$, where the mixture distribution is non-identifiable.
In particular, there is a mixture on topology $T_1$ and also a mixture on $T_3$ which
generate identical distributions.  Hence, it is impossible to determine the correct
topology in the worst case.  It turns out that this example does not extend to models
such as JC and K2.  In fact, all mixtures in JC and K2 models are identifiable.  This
follows from our following duality theorem which distinguishes which models
have non-identifiable mixture distributions, or have an easy method to determine
the common topology in the mixture.

We prove, that for any model which is parameterized by an open set, either
there exists a linear test (which is a strictly separating hyperplane as defined shortly), or the
model has {\em non-identifiable}  mixture distributions in the following sense.
Does there exist a tree $T$, a collection $\vec{P_1},\dots,\vec{P_k}$,
and distribution $p_1,\dots,p_k$ , such that there
is another tree $T'\neq T$, a  collection $\vec{P'_1},\dots,\vec{P''_k}$
and a distribution $p'_1,\dots,p'_k$
where:
\[
   \sum_{i=1}^k p_i\mu_{T,\vec{P_i}} = \sum_{i=1}^k p'_i\mu_{T',\vec{P'_i}}
   \]
   Thus in this case it is impossible to distinguish these two distributions.  Hence, we
   can not even infer which of the topologies $T$ or $T'$ is correct.
If the above holds, we say the model has {\em non-identifiable mixture distributions}.

In contrast, when there is no non-identifiable mixture distribution we can use the following notion of a
linear test to reconstruct the topology.
A {\em linear test} is a hyperplane strictly separating distributions arising from two different 4 leaf trees
(by symmetry the test can be used to distinguish between the 3 possible 4 leaf trees).
It suffices to consider trees with 4 leaves, since the full topology can be inferred from
all 4 leaf subtrees (Bandelt and Dress \cite{Bandelt}).

Our duality theorem uses a geometric viewpoint (see Kim \cite{Kim} for a nice
introduction to a geometric approach).
Every mixture distribution $\mu$ on a 4-leaf tree $T$
defines a point $z\in\real^N$ where $N=|\Omega|^4$.
For example, for the CFN model, we have
$z=(z_1,\dots,z_{2^4})$ and $z_1=\mu(0000), z_2=\mu(0001), z_3=\mu(0010),
\dots, z_{2^4}=\mu(1111)$.
Let $C_i$ denote the set of
points corresponding to distributions $\mu(T_i,\vec{P})$ for the 4-leaf tree $T_i$, $i=1,2,3$.
A linear test is a hyperplane which strictly separates the sets for a pair of trees.

\BDEF Consider the 4-leaf trees $T_2$ and $T_3$.
A {\em linear test} is a vector $t\in\real^{|\Omega|^4}$ such that
$t^T\mu_2>0$ for any mixture distribution $\mu_2$ arising from $T_2$
and $t^T\mu_3< 0$ for any mixture distribution $\mu_3$ arising from
$T_3$. \EDEF

There is
nothing special about $T_2$ and $T_3$ - we can distinguish between
mixtures arising from any two $4$ leaf trees, e.\,g., if $t$ is a
test then $t^{(1\,3)}$ distinguishes the mixtures from $T_1$ and
the mixtures from $T_2$, where $(1\,3)$ swaps the labels for
leaves $1$ and $3$.  More precisely, for all $(a_1,a_2,a_3,a_4)\in|\Omega|^4$,
\begin{equation}
\label{testswap}
t^{(1\,3)}_{a_1,a_2,a_3,a_4} = s_{a_4,a_2,a_3,a_1}
\end{equation}

\begin{theorem}\label{thm:dual}
For any model whose set ${\cal M}$ of transition matrices is parameterized by
an open set (of multilinear polynomials),
exactly one of the following holds:
\begin{itemize}
\item there exist non-identifiable mixture distributions, or
\item there exists a linear test.
\end{itemize}
\end{theorem}

For the JC and K2 models, the existence of a linear test follows immediately
from Lake's linear invariants \cite{Lake}.  Hence, our duality theorem implies that
there are no non-identifiable mixture distributions in this model.
In contrast for the K3 model, we prove there is no linear test, hence there is an
non-identifiable mixture distribution. We also prove that in the K3 model in the
common rate matrix framework, there is a non-identifiable mixture distribution.

To summarize, we show the following:
\begin{theorem}\hfil
\begin{enumerate}
\item  \label{thm:main222}
In the CFN model, there is an ambiguous mixture of size 2.
\item
In the JC and K2 model, there are no ambiguous mixtures.
\item In the K3 model there exists a non-identifiable mixture distribution
(even in the common rate matrix framework).
\end{enumerate}
\end{theorem}

Steel, Sz\'{e}kely and Hendy \cite{SSH} previously proved the
existence of a non-identifiable mixture distribution in the CFN model, but their
proof was non-constructive and gave no bound on the size of the mixture.
Their result had the more appealing
feature that the trees in the mixture were scalings of each other.

Allman and Rhodes \cite{AllmanRhodes} recently proved
identifiability of the topology for certain classes of mixture
distributions using invariants (not necessarily linear).
Rogers \cite{Rogers} proved that the topology is identifiable in
the general time-reversible model when the rates vary
according to what is known as the invariable sites plus gamma distribution model.
Much of the current work on invariants uses ideas from algebraic
geometry, whereas our notion of a linear test is natural from the perspective
of convex programming duality.

Note, that even in models that do not have non-identifability between different topologies,
there is non-identifiability within the topology.
An interesting example was
shown by Evans and Warnow \cite{EvansWarnow}.

\subsection{Outline of Paper}

We prove, in Section \ref{sec:duality}, Theorem \ref{thm:dual} that a phylogenetic model has
a non-identifiable mixture distribution or a linear test.  We then detail
Lake's linear invariants in Section \ref{sec:tests}, and conclude the existence of a linear
test for the JC and K2 models.
In Sections \ref{sec:ambiguity} and \ref{sec:K3} we prove that there are non-identifiable mixtures in
the CFN and K3 models, respectively.  We also present a linear test for a restricted version of the CFN model
in Section \ref{sec:CFNtest}.  We prove the maximum likelihood results
stated in Theorem \ref{thm:MLEbad} in Section \ref{sec:maxlik}.
The maximum likelihood results require several technical tools which are also proved in
Section \ref{sec:maxlik}.  The MCMC results
are then stated formally and proved in Section \ref{sec:mcmc}.

\section{Preliminaries}
\label{sec:notation}

Let the permutation group $S_4$ act on
the 4-leaf trees $\{T_1,T_2,T_3\}$ by renaming the leaves. For example $(14)\in
S_4$ swaps $T_2$ and $T_3$, and fixes $T_1$.
For $\pi\in S_n$, we let $T^\pi$ denote tree $T$ permuted by $\pi$.
It is easily checked that the following group $K$ (Klein group)
fixes every $T_i$:
\begin{equation}\label{klein}
K=\{(),(1\,2)(3\,4),(1\,3)(2\,4),(1\,4)(2\,3)\}\leq S_4.
\end{equation}
Note that $K\unlhd S_4$, i.\,e., $K$ is a normal subgroup of $S_4$.

For weighted trees we let $S_4$ act on $(T,\vec{P})$ by changing
the labels of the leaves but leaving the weights of the edges
untouched.
Let $\pi\in S_n$ and let $T'=T^\pi$. Note that the distribution
$\mu_{T',w}$ is just a permutation of the distribution
$\mu_{T,w}$:
\begin{equation}\label{e:perm}
\mu_{T',\vec{P}}(a_{\pi_1},\dots,a_{\pi_n})=\mu_{T,\vec{P}}(a_1,\dots,a_n).
\end{equation}
The actions on weighted trees and on
distributions are compatible: $$ \mu_{(T,\vec{P})^\pi} =
(\mu_{T,\vec{P}})^\pi.$$

\section{Duality Theorem}
\label{sec:ambiguity-or-test}
\label{sec:duality}

In this section we prove the
duality theorem (i.e., Theorem \ref{thm:dual}).

Our assumption that the transition matrices of the model are
parameterized by an open set implies the following observation.
\BOBS\label{obs:multisecond}
For models parameterized by an open set, the coordinates of
$\mu'_{T,w}$ are multi-linear polynomials in
the parameters.
\EOBS

\begin{figure}[htb]
\begin{center}
\includegraphics[type=eps,ext=.eps,read=.eps,height=.7in]{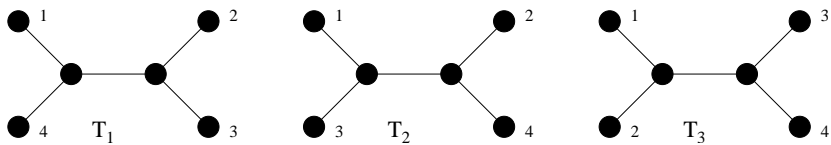}
\caption{All leaf-labeled binary trees with $n=4$ leaves.}\label{trees}
\end{center}
\end{figure}

We now state a classical result that allows one to
reduce the reconstruction problem to trees with 4 leaves.
Note there are three distinct leaf-labeled binary trees with $4$
leaves. We will call them $T_1,T_2,T_3$ see Figure \ref{trees}.

For a tree $T$ and a set $S$ of leaves, let
$T|_S$ denote the induced subgraph of $T$ on $S$ where
internal vertices of degree $2$ are removed.

\begin{theorem}[Bandelt and Dress \cite{Bandelt}]
For distinct
leaf-labeled binary trees $T$ and $T'$ there exist a set $S$ of $4$ leaves where
$T|_S\neq T'|_S$.   Hence, the set of induced subgraphs on all 4-tuples of
leaves determines a tree.
\end{theorem}

The above theorem also simplifies the search for
non-identifiable mixture distributions.
\begin{corollary}
If there exists a non-identifiable mixture distribution
then there exists a non-identifiable mixture distribution on trees with $4$ leaves.
\end{corollary}

Recall from the Introduction,
the mixtures arising from $T_i$ form a convex set in the space of
joint distributions on leaf labelings. A test is a hyperplane {\em
strictly} separating the mixtures arising from $T_2$ and the
mixtures arising from $T_3$. For general disjoint convex sets a
strictly separating hyperplane need not exist (e.\,g., take
$C_1=\{(0,0)\}$ and $C_2=\{(0,y)\,|\,y>0\}\cup\{(x,y)\,|\,x>0\}$).
The sets of mixtures are special - they are convex hulls of images
of open sets under a multi-linear polynomial map.

\BLEM\label{lem:dua} Let
$p_1(x_1,\dots,x_n),\dots,p_m(x_1,\dots,x_n)$ be multi-linear
polynomials in $x_1,\dots,x_n$. Let $p=(p_1,\dots,p_m)$. Let
$D\in\real^n$ be an open set. Let $C$ be the convex hull of
$\{p(x)\,|\,x\in D\}$. Assume that $0\not\in C$. There exists
$s\in\real^m$ such that $s^Tp(x)>0$ for all $x\in D$. \ELEM

\BPRF
Suppose the polynomial $p_i$ is a linear combination of the other
polynomials, i.e., $p_i(x)=\sum_{j\neq i} c_jp_j(x)$.
Let $p'=(p_1,\dots,p_{i-1},p_{i+1},\dots,p_m)$.
Let $C'$ be the convex hull of $\{p'(x)\,|\,x\in D\}$.
Then
\[
(y_1,\dots,y_{i-1},y_{i+1},\dots,y_m)\mapsto
\left(y_1,\dots,y_{i-1},\sum_{j\neq i}c_jy_j,y_{i+1},\dots,y_m\right)
\]
is a bijection between points in $C'$ and $C$.
Note, $0\not\in C'$.
There exists a strictly separating hyperplane between $0$ and $C$
(i.e., there exists $s\in\real^m$ such that $s^Tp(x)>0$ for all $x\in D$)
if and only if there exists a strictly separating hyperplane between $0$ and
$C'$ (i.e., there exists $s'\in\real^m$ such that $s'^Tp'(x)>0$ for all $x\in D$).
Hence, without loss of generality,
we can assume that the polynomials $p_1,\dots,p_m$
are linearly independent.

Since $C$ is convex and $0\not\in C$, by the
separating hyperplane theorem, there exists $s\neq 0$ such that
$s^Tp(x)\geq 0$ for all $x\in D$. If $s^Tp(x)>0$ for all $x\in D$
we are done.

Now suppose $s^Tp(a)=0$ for some $a\in D$. If $a\neq 0$, then
by translating $D$ by $-a$ and changing the polynomials appropriately
(namely, $p_j(x):=p_j(x+a)$),
without loss of generality, we can assume $a=0$.

Let $r(x_1,\dots,x_n)=s^Tp(x_1,\dots,x_n)$. Note that $r$
is multi-linear and $r\neq 0$ because $p_1,\dots,p_m$ are linearly
independent. Since $a=0$, we have $r(0)=0$ and hence $r$ has no constant
monomial.


Let $w$ be the monomial of lowest total degree which has a non-zero
coefficient in $r$.  Consider $y=(y_1,\dots,y_m)$ where
$y_j\neq 0$ for $x_j$ which occur in $w$ and
$y_j=0$ for all other $x_j$.  Then,
$r(y)=w(j)$ since
there are no monomials of smaller degree, and
any other monomials contain some $y_j$ which is $0$.
Hence by choosing $y$ sufficiently close to $0$, we have
$y\in D$ (since $D$ is open) and $r(y)<0$ (by choosing
an appropriate direction for $y$).  This contradicts the
assumption that $s$ is a separating hyperplane.
Hence $r=0$ which is a
contradiction with the linear independence of
the polynomials $p_1,\dots,p_m$.
 \EPRF

We now prove our duality theorem.

\begin{proof}[Proof of Theorem \ref{thm:dual}]
Clearly there cannot exist both a non-identifiable mixture and a
linear test. Let $C_i$ be the convex set of mixtures arising from $T_i$
(for $i=2,3$).
Assume that $C_2\cap C_3=\emptyset$, i.\,e., there
is no non-identifiable mixture in ${\cal M}$. Let $C=C_2-C_3$. Note that
$C$ is convex, $0\not\in C$, and $C$ is the convex hull of an
image of open sets under multi-linear polynomial maps (by
Observation \ref{obs:multisecond}).
 By Lemma~\ref{lem:dua}
there exists $s\neq 0$ such that $s^Tx>0$ for all $x\in C$. Let
$t=s-s^{(1\,4)}$ (where $s^{(1\,4)}$ is defined as in \eqref{testswap}).
Let $\mu_2\in C_2$ and let
$\mu_3=\mu_2^{(1\,4)}$ where $\mu_2^{(1\, 4)}$ is defined analogously to
$s^{(1\, 4)}$. Then
$$t^T\mu_2=(s-s^{(1\,4)})^T\mu_2=s^T\mu_2-s^T\mu_3=s^T(\mu_2-\mu_3)>0.$$
Similarly for $\mu_3\in C_3$ we have $t^T\mu_3<0$ and hence $t$ is
a test. \EPRF

\section{Simplifying the search for a linear test}

The CFN, JC, K2 and K3 models all have a natural group-theoretic structure.
We show some key properties of linear tests utilizing this structure.
These properties will simplify
proofs of the existence of linear tests in JC and K2 (and restricted CFN) models
and will also be used in the proof of the non-existence of linear tests in K3 model.
Our main objective is to use symmetry inherent in the phylogeny setting
to drastically reduce the dimension of the space
of linear tests.

Symmetric phylogeny models have a group of symmetries $G\leq S_{\Omega}$
($G$ is the intersection of the automorphism groups of the
weighted graphs corresponding to the matrices in ${\cal M}$). The
probability of a vertex labeling of $T$ does not change if the
labels of the vertices are permuted by an element of $G$. Thus the
elements of $\Omega^n$ which are in the same orbit of the action
of $G$ on $\Omega^n$ have the same probability in any distribution
arising from the model.

Let $O'$ be the orbits of $\Omega^4$ under the action of $G$.
 Let
$O$ be the orbits of $O'$ under the action of $K$. Note that the
action of $(1\,4)$ on $O$ is well defined (because $K$ is a normal
subgroup of $S_4$). For each pair $o_1,o_2\in O$ that are swapped by
$(1\,4)$ let
\begin{equation}\label{eq:test}
\ell_{o_1,o_2}(\mu)=\sum_{a\in o_1}\mu(a)-\sum_{a\in o_2}\mu(a).
\end{equation}

\BLEM\label{lem:tes}
Suppose that ${\cal M}$ has a linear test $s$. Then ${\cal M}$ has a linear test
$t$ which is a linear combination of the $\ell_{o_1,o_2}$.\ELEM

\BPRF
Let $s$ be a linear test. Let
$$
t'=\sum_{g\in K} s^g\quad \mbox{and}\quad t=t'-t'^{(1\,4)}.
$$
Let $\mu_2$ be a mixture arising from $T_2$. For any $g\in K$ the mixture
$\mu_2^g$ arises from $T_2$ and hence
$$
(t')^T\mu_2=\sum_{g\in K} (s^g)^T\mu_2=
\sum_{g\in K} s^T(\mu^g_2)>0.$$
Similarly $(t')^T\mu_3<0$ for $\mu_3$ arising from $T_3$ and hence
$t'$ is a linear test.

Now we show that $t$ is a linear test as well. Let $\mu_2$ arise from
$T_2$. Note that $\mu_3=\mu_2^{(1\,4)}$ arises from $T_3$ and
hence $$t^T\mu_2=(t'-t'^{(14)})^T\mu_2=(t')^T\mu_2-(t')^T\mu_3>0.$$
Similarly $t(\mu_3)<0$ for $\mu_3$ arising from $T_3$ and hence
$t$ is a linear test.

Note that $t$ is zero on orbits fixed by $(1\,4)$. On orbits
$o_1,o_2$ swapped by $(1\,4)$ we have that $t$ has opposite value
(i.\,e., $a$ on $o_1$, and $-a$ on $o_2$ for some $a$). Hence
$t$ is a linear combination of the $\ell_{o_1,o_2}$.
\EPRF

\subsection{A simple condition for a linear test}

For the later proofs it will be convenient to label the edges by
matrices which are not allowed by the phylogenetic models.
For example the identity matrix $I$ (which corresponds
to zero length edge) is an invalid transition matrix, i.e., $I\not\in\mathcal{M}$,
for the models considered in this paper.

The definition (\ref{eq:defp}) is continuous in the entries of the
matrices and hence for a weighting by matrices in ${\rm cl}({\cal
M})$ (the closure of ${\cal M}$) the generated distribution is
arbitrarily close to a distribution generated from the model.

\begin{observation}
\label{closure}
A linear test for $\mathcal{M}$ (which is a strictly separating hyperplane for ${\cal M}$)
is a separating hyperplane for ${\rm cl}({\cal M})$.
\end{observation}

The above observation follows
from the fact that if a continuous function
$f:cl(A)\ra\real$ is positive on some set $A$ then it is non-negative on ${\rm cl}(A)$.

Suppose that the identity matrix $I\in {\rm cl}({\cal M})$. Let
$\mu$ arise from $T_2$ with weights such that the internal edge
has weight $I$. Then $\mu_2$ arises also from $T_3$ with the
same weights. A linear test has to be positive for mixtures form $T_2$
and negative for mixtures from $T_3$. Hence we have:

\BOBS\label{obs:zero}
Let $\mu$ arise from $T_2$ with weights such that the internal
edge has transition matrix $I$. Let $t$ be a linear test. Then $t^T\mu=0$.
\EOBS

\section{Linear tests for JC and K2}
\label{sec:tests}

In this section we show a linear test for JC and K2 models.  In fact we
show that the linear invariants introduced by Lake \cite{Lake} are linear
tests.  We expect that this fact is already known, but we include the proof for
completeness and since it almost elementary given the preliminaries
from the previous section.

\subsection{A linear test for the Jukes-Cantor model}

To simplify many of the upcoming expressions throughout the following section,
we center the
transition matrix for the Jukes-Cantor (JC) model around its stationary distribution
in the following manner.
Recall the JC model has $\Omega_{{\rm JC}}=\{0,1,2,3\}$
and
its semigroup ${\cal M}_{\rm JC}$ consists of matrices
$$
M_{\rm JC}(\xhat)=\frac{1}{4}E
+\left(
\begin{array}{cccc}
3\xhat & -\xhat & -\xhat & -\xhat \\
-\xhat& 3\xhat& -\xhat& -\xhat\\
-\xhat& -\xhat& 3\xhat& -\xhat\\
-\xhat& -\xhat& -\xhat& 3\xhat
\end{array}
\right),
$$
where $E$ is the all ones matrix (i.e., $E(i,j) =1$ for all $0\leq i,j<|\Omega|$)
 and $0<\xhat<1/4$.

We refer to $\xhat$ as the {\em centered edge weight}.
Thus, a centered edge weight of $\xhat=1/4$ (which is not valid) means both
endpoints have the same assignment.  Whereas $\xhat=0$ (also not valid)
means the endpoints are independent.

The group of symmetries of $\Omega_{\rm JC}$ is $G_{{\rm
JC}}=S_4$. There are $15$ orbits in $\Omega_{\rm JC}^4$ under the
action of $G_{\rm JC}$ (each orbit has a representative in which
$i$ appears before $j$ for any $i<j$). The action of $K$ further
decreases the number of orbits to $9$. Here we list the $9$ orbits
and indicate which orbits are swapped by $(1\,4)$:
\begin{equation}\label{jco}
\begin{small}
\begin{array}{l}0000\end{array},
\begin{array}{l}0110\end{array},
\begin{array}{l}0123\end{array},
\begin{array}{l}0112\\0120\end{array},
\begin{array}{l}0111\\0100\\0010\\0001\end{array},
\begin{array}{l}0011\end{array}\leftrightarrow\begin{array}{l}0101\end{array},
\begin{array}{l}0122\\0012\end{array}\leftrightarrow\begin{array}{l}0121\\0102\end{array}.
\end{small}
\end{equation}
By Lemma~\ref{lem:tes} every linear test in the JC model is a linear combination of
\begin{eqnarray}
\nonumber
t_1&=&\mu(0011)-\mu(0101),\ \mbox{and}
\\
\label{testcj}
t_2&=&\mu(0122)-\mu(0121)-\mu(0102)+\mu(0012).
\end{eqnarray}

We will show that $t_2-t_1$ is a linear test and that there
exist no other linear tests (i.\,e., all linear tests are multiples of
$t_2-t_1$).

\BLEM\label{lcj}
Let $\mu$ be a single-tree mixture arising from a tree $T$ on $4$ leaves.
Let $t$ be defined by
\begin{eqnarray}
\nonumber
t^T\mu=t_2-t_1&=&\mu(0122)-\mu(0121)+\mu(0101)
\\ && \ \ -\mu(0011)+\mu(0012)-\mu(0102).
\label{ephi}
\end{eqnarray}
Let $\mu_i$ arise from $T_i$, for $i=1,2,3$. We have
$$t^T\mu_1=0,\quad t^T\mu_2>0,\ \mbox{and}\ t^T\mu_3<0.$$
In particular $t$ is a linear test.
\ELEM

 \BPRF
 Label the 4 leaves as $v_1,\dots,v_4$, and let $\xhat_1,\dots,\xhat_4$
denote the centered edge weight of the edge incident to the respective leaf.
Let $\xhat_5$ denote the centered edge weight of the internal edge.

Let $\mu_j$ arise from $T_j$ with centered edge weights $\xhat_1,\dots,\xhat_5$,
$j\in\{1,2,3\}$. Let $\Phi_j$ be the multi-linear polynomial
$t^T\mu_j$. If $\xhat_4=0$ then $\mu_j$
does not depend on the label of
$v_4$ and hence, for all $a\in\Omega_{JC}$,
 $$
\Phi_j=
\mu_j(012a)-\mu_j(012a)+\mu_j(010a)-\mu_j(001a)+\mu_j(001a)-\mu_j(010a)=0.
 $$
Thus $\xhat_4$ divides $\Phi_j$. The $t_i$ are invariant under the
action of $K$ (which is transitive on $1,2,3,4$) and hence
$\Phi_j$ is invariant under the action of $K$. Hence $\xhat_i$ divides
$\Phi_j$ for $i=1,\dots,4$.  We have
$$\Phi_j=\xhat_1\xhat_2\xhat_3\xhat_4\ell(\xhat_5), $$
where $\ell(\xhat_5)$ is a linear
polynomial in $\xhat_5$.

Let $\mu'_1$ arise from $T_1$ with $\xhat_1=\dots=\xhat_4=1/4$.
 In leaf-labelings with non-zero probability in $\mu'_1$ the labels of
$v_1,v_4$ agree and the labels of $v_2,v_3$ agree. None of the
leaf-labelings in (\ref{ephi}) satisfy this requirement and hence
$\Phi_1=0$ if $\xhat_1=\dots=\xhat_4=1/4$. Hence $\ell(\xhat_5)$ is the zero
polynomial and $\Phi_1$ is the zero polynomial as well.

Now we consider $T_2$. If $\xhat_5=1/4$ then, by
Observation~\ref{obs:zero}, $\Phi_2=0$. Thus $1/4$ is a root of
$\ell$ and hence $\Phi_2=\alpha\cdot \xhat_1\xhat_2\xhat_3\xhat_4(1/4-\xhat_5)$.
We plug in
$\xhat_5=0$ and $\xhat_1=\xhat_2=\xhat_3=\xhat_4=1/4$ to determine $\alpha$. Let
$\mu_2'$ be the distribution generated by these weights. The
leaf-labelings for which $\mu'_2$ is non-zero must have the same
label for $v_1,v_3$ and the same label for $v_2,v_4$. Thus
$\Phi_2=\mu'_2(0101)=1/16$ and hence $\alpha=64$. We have
$$\Phi_2=64 \xhat_1\xhat_2\xhat_3\xhat_4(1-4\xhat_5).$$
Note that $\Phi_2$ is always positive. The action of $(1\,4)$
switches the signs of the $t_i$ and hence $\Phi_3=-\Phi_2$. Thus
$\Phi_3$ is always negative.
\EPRF

We now show uniqueness of the above linear test, i.e., any other linear test
is a multiple of $t_2-t_1$.

\BLEM
Any linear test in the JC model is a multiple of (\ref{ephi}).
\ELEM

\BPRF
Let $t=\alpha_1 t_1+\alpha_2 t_2$ be a linear test. Let $\mu_1$
be the distribution generated by centered weights $\xhat_2=\xhat_4=\xhat_5=1/4$ and
$\xhat_1=\xhat_3=0$ on $T_2$. By Observation~\ref{obs:zero} we must
have $t^T\mu_1=0$. Note that
\begin{equation*}\label{eq:p1}
\mu_1(a_1a_2a_3a_4)=\Big\{
\begin{array}{ll}
1/64 & \mbox{if}\ a_2=a_4,\\
0 & \mbox{otherwise}.
\end{array}
\end{equation*}
Hence
\begin{equation*}\label{eq:q1}
t^T\mu_1=-\alpha_1\mu_1(0101)-\alpha_2\mu(0121)=-1/64(\alpha_1+\alpha_2)=0.
\end{equation*}
Thus $\alpha_1=-\alpha_2$ and hence $t$ is a multiple of (\ref{ephi}).
\EPRF

\subsection{A linear test for Kimura's 2-parameter model}
\label{sec:K2}

Mutations between two purines ({\A} and \G) or between two pyrimidines ({\C} or \T)
are more likely than mutations between a purine and a pyrimidine.
Kimura's 2-parameter model (K2) tries to model this fact.

We once again center the transition matrices to simplify the calculations.
The K2 model has $\Omega_{{\rm K2}}=\{0,1,2,3\}$ and
its semigroup ${\cal M}_{\rm K2}$ consists of matrices
$$
M_{\rm K2}(\xhat,\yhat)=\frac{1}{4}E
+\left(
\begin{array}{cccc}
\xhat+2\yhat & -\xhat & -\yhat & -\yhat \\
-\xhat & \xhat+2\yhat & -\yhat & -\yhat \\
-\yhat & -\yhat & \xhat+2\yhat & -\xhat \\
-\yhat & -\yhat & -\xhat & \xhat+2\yhat
\end{array}
\right), $$
with $\xhat\leq \yhat<1/4$ and $\xhat+\yhat>0$.
See Felsenstein \cite{F:book} for closed form of the transition matrices of the model
in terms of the times $t_e$ and rate matrices $R_e$.  One can then derive
the equivalence of the conditions there with the
conditions $\xhat\leq \yhat<1/4, \xhat+\yhat>0$.

Note, $\xhat$ can be negative, and hence certain transitions can have
probability $>1/4$ but are always $<1/2$.
Observe that $M_{\rm K2}(\xhat,\xhat)=M_{\rm JC}(\xhat)$,
i.e., the JC model is a special case of the K2 model.

The group of symmetries is $G_{{\rm K2}}=\<(01),(02)(13)\>$ (it
has $8$ elements). There are $36$ orbits in $\Omega^4$ under the
action of $G_{\rm K2}$ (each orbit has a representative in which
$0$ appears first and $2$ appears before $3$). The action of $K$
further decreases the number of orbits to $18$. The following
orbits are fixed by $(1\,4)$: $$
\begin{small}
\begin{array}{l}0000\end{array},
\begin{array}{l}0110\end{array},
\begin{array}{l}0220\end{array},
\begin{array}{l}0231\end{array},
\begin{array}{l}0221\\0230\end{array},
\begin{array}{l}0111\\0100\\0010\\0001\end{array},
\begin{array}{l}0222\\0200\\0020\\0002\end{array},
\begin{array}{l}0223\\0210\\0120\\0112\end{array}
\end{small}.
$$
The following orbits are swapped by $(1\,4)$:
$$
\begin{small}
0011\leftrightarrow 0101,
0022\leftrightarrow 0202,
0123\leftrightarrow 0213,
\begin{array}{l}0122\\0023\end{array}\leftrightarrow\begin{array}{l}0212\\0203\end{array},
\begin{array}{l}0233\\0211\\0021\\0012\end{array}\leftrightarrow
\begin{array}{l}0232\\0121\\0201\\0102\end{array}.
\end{small}
$$
By Lemma \ref{lem:tes} any linear test for the K2 model is a linear combination of
\begin{eqnarray*}
t_1&=&\mu(0011)-\mu(0101),\\
t_2&=&\mu(0233)+\mu(0211)-\mu(0232)-\mu(0201)
\\ & & \ \ -\mu(0121)+\mu(0021)-\mu(0102)+\mu(0012),\\
t_3&=&\mu(0022)-\mu(0202),\\
t_4&=&\mu(0122)+\mu(0023)-\mu(0212)-\mu(0203),\\
 t_5&=&\mu(0123)-\mu(0213)
\end{eqnarray*}

\BLEM \label{lk2}
Let $\mu$ be a single-tree mixture arising from a tree $T$ on $4$ leaves.
Let $t$ be defined by
\begin{eqnarray}
\nonumber
t^T\mu&=&\mu(0122)-\mu(0212)+\mu(0202)-\mu(0022)+
\\ && \ \ \mu(0023)-\mu(0203)+\mu(0213) -\mu(0123).
\label{ephi2}
\end{eqnarray}
Let $\mu_i$ arise from $T_i$, for $i=1,2,3$. We have
$$t^T\mu_1=0,\quad t^T\mu_2>0,\ \mbox{and}\ t^T\mu_3<0.$$
In particular $t$ is a linear test. \ELEM

\BPRF Let $T=T_j$ for some $j\in\{1,2,3\}$. Let the transition matrix
of the edge incident to leaf $v_i$ be $M_{\rm K2}(\xhat_i,\yhat_i)$, and
the internal edge has $M_{\rm K2}(\xhat_5,\yhat_5)$.
Let $\mu_j$ be the generated
distribution, and let $\Phi_j$ be the multi-linear polynomial
$t^T\mu_j$.

If $\yhat_4=-\xhat_4$ then the matrix on the edge incident to leaf $v_4$
has the last two
columns the same.  Hence roughly speaking this edge
forgets the distinction between labels $2$ and $3$, and therefore,
in \eqref{ephi2}, we can do the following replacements:
\begin{eqnarray*}
  0122 & \rightarrow & 0123 \\
  0202 & \rightarrow & 0203 \\
0022 &\rightarrow & 0023 \\
0213 &\rightarrow & 0212,
  \end{eqnarray*}
 and we obtain,
\begin{equation}\label{e4la}
\Phi_j=0.
\end{equation}

Thus $\xhat_4-\yhat_4$ divides $\Phi_j$. Since $\Phi_j$ is invariant under
the action of $K$ we have that $\xhat_i-\yhat_i$ divides $\Phi_j$ for
$i=1,\dots,4$ and hence
\begin{equation}
\label{eq:dividesj}
\Phi_j=(\xhat_1-\yhat_1)\dots(\xhat_4-\yhat_4)\ell_j(\xhat_5,\yhat_5),
\end{equation} where
$\ell_j(\xhat_5,\yhat_5)$ is linear in $\xhat_5$ and $\yhat_5$.

Now let $\xhat_i=\yhat_i=1/4$ for $i=1,\dots,4$. The label of the internal
vertices $v_j$ for $j=5,6$ must agree with the labels of neighboring leaves and
hence
\begin{equation}\label{no13}
\Phi_j=\mu_j(0202)-\mu_j(0022).
\end{equation}
Now plugging $j=1$ into \eqref{eq:dividesj} and \eqref{no13},
for this setting of $\xhat_i,\yhat_i$,
we have
\begin{equation}\label{emu1}
\Phi_{1}=0
\end{equation}
By plugging $j=2,3$ into \eqref{eq:dividesj} and \eqref{no13} we have
\begin{equation}\label{emu2}
\Phi_{2}=-\Phi_{3}=(\xhat_1-\yhat_1)(\xhat_2-\yhat_2)(\xhat_3-\yhat_3)(\xhat_4-\yhat_4)(1-4\yhat_5).
\end{equation}
Note that $\xhat_i-\yhat_i<0$ and $\yhat_i<1/4$ and hence (\ref{emu2}) is
always positive. Linearity of the test $\mu\mapsto t^T\mu$
implies
that $t^T\mu$ is positive for any mixture generated from $T_2$
and negative for any  mixture generated from $T_3$. \EPRF

\BLEM\label{k2unique}
Any linear test in the K2 model is a multiple of (\ref{ephi2}).
\ELEM

\BPRF
Let $t=\alpha_1 t_1+\dots+\alpha_5 t_5$ be a linear test. A linear test in the
K2 model must work for JC model as well. Applying symmetries
$G_{\rm JC}$ we obtain

\begin{equation}\label{k2jkfuj}
\begin{split}
t=\alpha_1(\mu(0011)-\mu(0101))+2\alpha_2(\mu(0122)-\mu(0121)-\mu(0102)+\mu(0012))\\
+\alpha_3(\mu(0011)-\mu(0101))+\alpha_4(\mu(0122)+\mu(0012)-\mu(0121)-\mu(0102)).
\end{split}
\end{equation}
Comparing (\ref{k2jkfuj}) with (\ref{ephi}) we obtain
\begin{equation}\label{k2jc}
\alpha_1+\alpha_3=-1\ \mbox{and}\ \alpha_4+2\alpha_2=1.
\end{equation}

Let $\mu_1$ arise from $T_2$ with centered
weights $(\xhat_2,\yhat_2)=(\xhat_4,\yhat_4)=(\xhat_5,\yhat_5)=(1/4,1/4)$,
$(\xhat_1,\yhat_1)=(1/4,0)$, and $(\xhat_3,\yhat_3)=(\xhat,\yhat)$.
From observation~\ref{obs:zero} it
follows that $\Phi_{\mu_1}=0$. The leaf-labelings with non-zero
probability must give the same label to $v_2$ and $v_4$, and the
labels of $v_1$ and $v_2$ must either be both in $\{0,1\}$ or both
in $\{2,3\}$. The only such leaf-labelings involved in
$t_1,\dots,t_5$ are $0101,0121$. Thus
\begin{equation}\label{eq:q1b}
t^T\mu_1=-\alpha_1\mu_1(0101)-\alpha_2\mu_1(0121)=-\frac{1}{16}(\alpha_1\xhat+\alpha_2\yhat)=0.
\end{equation}
Thus $\alpha_1=\alpha_2=0$ and from (\ref{k2jc}) we get
$\alpha_3=-1$ and $\alpha_4=1$.

Let $\mu_2$ be generated from $T_2$ with centered weights
$(\xhat_4,\yhat_4)=(\xhat_5,\yhat_5)=(1/4,1/4)$,
$(\xhat_1,\yhat_2)=(\xhat_3,\yhat_3)=(0,0)$, and
$(\xhat_2,\yhat_2)=(1/4,0)$. In
leaf-labelings with non-zero probability the labels of $v_2$ and
$v_4$ are either both in $\{0,1\}$ or both in $\{2,3\}$. The only
such leaf labelings in $t_3,t_4,t_5$ are 0202,0213,0212, and 0203.
Hence
$$t^T\mu_3=\mu_3(0202)-\mu_3(0212)-\mu_3(0203)-\alpha_5\mu_3(0213)=
\frac{1}{256}(3-3-1-\alpha_5)=0.$$
Thus $\alpha_5=-1$ and all the
$\alpha_i$ are determined. Hence the linear test is unique (up to scalar
multiplication). \EPRF

\section{Non-identifiability and linear tests in CFN}
\label{sec:ambiguity}

In this section we consider the CFN model.
We first prove there is no linear test, and then we
present a non-identifiable mixture distribution.
We then show that there is
a linear test for the CFN model when the edge probabilities are restricted to
some interval.

\subsection{No linear test for CFN}

Again, when considering linear tests
we look at the model with its transition matrix centered
around its stationary distribution.  The CFN model has $\Omega_{{\rm CFN}}=\{0,1\}$ and
its semigroup ${\cal M}_{\rm CFN}$ consists of matrices
$$
M_{\rm CFN}(\xhat)=\frac{1}{2}E
+\left(
\begin{array}{cc}
\xhat & -\xhat \\
-\xhat & \xhat
\end{array} \right)
$$
with $0<\xhat<1/2$.

In the CFN model,
note that the roles of $0$ and $1$ are symmetric, i.\,e.,
\begin{equation}\label{e:comp}
\mu_{T,\vec{P}}(a_1,\dots,a_n)=\mu_{T,\vec{P}}(1-a_1,\dots,1-a_n).
\end{equation}
Hence the group of
symmetries of $\Omega_{{\rm CFN}}$ is $G_{{\rm CFN}}=\langle (0
1)\rangle=\intg/(2\intg)$. There are $8$ orbits of the action of $G_{\rm
CFN}$ on $\Omega_{\rm CFN}^4$ (one can choose a representative for
each orbit to have the first coordinate $0$). The action of $K$
further reduces the number of orbits to $5$. The action of
$(1\,4)$ swaps two of the orbits and keeps $3$ of the orbits
fixed:
$$
\begin{small}
\begin{array}{l}0000\end{array},
\begin{array}{l}0110\end{array},
\begin{array}{l}0111\\0100\\0010\\0001\end{array},
\begin{array}{l}0011\end{array}\leftrightarrow\begin{array}{l}0101\end{array}.
\end{small}
$$ By Lemma~\ref{lem:tes}, if there exists a linear test for CFN then (a
multiple of) $t_1=\mu(0011)-\mu(0101)$ is a linear test. Let $\mu$ arise from
$T_2$ with the edge incident to leaf $v_i$ labeled by $M_{\rm CFN}(\xhat_i)$, for
$i=1,\dots,4$, and the internal edge labeled by $M_{\rm CFN}(\xhat_5)$.
A short calculation yields
\begin{equation}\label{ecf}
\mu(0011)-\mu(0101)=\xhat_5(\xhat_1\xhat_3+\xhat_2\xhat_4)-\frac{\xhat_1\xhat_2+\xhat_3\xhat_4}{2}.
\end{equation}
Note that (\ref{ecf}) is negative if $\xhat_5$ is much smaller than
the other $\xhat_i$; and (\ref{ecf}) is positive if $\xhat_1,\xhat_3$ are much
smaller than the other $\xhat_i$. Thus $t_1$ is not a linear test and hence
there does not exist a linear test in the CFN model. By
Theorem~\ref{thm:dual} there exists a non-identifiable mixture. The next
result gives an explicit family of non-identifiable mixtures.

\subsection{Non-identifiable Mixture for CFN}

For each edge $e$ we will give the edge probability $0<w_e<1/2$, which is the probability
the endpoints receive different assignments (i.e., it is the off-diagonal entry
in the transition matrix.
For a 4-leaf tree $T$, we specify a set of transition matrices for
the edges by a 5-dimensional vector $\vector{P}=(w_1,w_2,w_3,w_4,w_5)$
where, for $1\le i\leq 4$, $w_i$ is the edge probability for the edge incident
to leaf labeled $i$, and $w_5$ is the edge probability for the internal edge.

\BPRO For $0< a,b< 1/2$ and $0<p\leq 1/2$, set
\begin{eqnarray*}
\vec{P_1} &=& \frac{1}{2}{\mathbf 1} - (a,b,a,b,c) \mbox{ and } \\
\vec{P_2} &=&\frac{1}{2}{\mathbf 1} - (b,a,b,a,d),
\end{eqnarray*}
where
${\mathbf 1}=(1,1,1,1,1)$ and
\begin{eqnarray*}
c&=&z/p, \\
d&=&z/(1-p), \mbox{ and } \\
z&=&\frac{ab}{2(a^2+b^2)}.
\end{eqnarray*}
Let $$\mu=p\mu_{T_3,\vec{P_1}}+(1-p)\mu_{T_3,\vec{P_2}}.$$
The distribution $\mu$ is invariant under $\pi=(14)$. Hence, $\mu$
is also generated by a mixture from $T^\pi$, a leaf-labeled tree different
from $T=T_3$. In particular, the following holds:
 $$\mu=p\mu_{T_2,\vec{P_1}}+(1-p)\mu_{T_2,\vec{P_2}}.$$
 Hence, whenever $c$ and $d$ satisfy $0<c,d<1/2$ then $\mu$ is
 in fact a distribution and there is non-identifiability.
 Note, for every $0<p\leq 1/2$, there exist $a$ and $b$ which define a
 non-identifiable mixture distribution.
\EPRO

\begin{proof}
Note that $\pi=(1\, 4)$ fixes leaf labels $0000, 0010, 0100, 0110$ and
swaps $0011$ with $0101$ and $0001$ with $0111$.

A short calculation yields
\begin{eqnarray*}
\mu(0011)-\mu(0101)&=&ab-(a^2+b^2)(pc+(1-p)d), \mbox{and}\\
\mu(0001)-\mu(0111)&=&(a^2-b^2)(pc-(1-p)d).
\end{eqnarray*}
which are both zero for our choice of $c$ and $d$. This implies that $\mu$ is invariant under the action of $(1\,4)$, and hence
non-identifiable.
\end{proof}

\subsection{Linear test for CFN with restricted weights}
\label{sec:CFNtest}

\BLEM Let $a \in (0,1/2)$. If the centered edge weight $\xhat$
for the CFN model is restricted to the interval $(a,\sqrt{a-a^2})$ then
there is a linear test.
\ELEM

\BPRF We will show that (\ref{ecf}) is positive if the $\xhat_i$ are
in the interval $(a,\sqrt{a-a^2})$. Let $b=\sqrt{a-a^2})$. Note
that $0<a<b<1/2$.

Since (\ref{ecf}) is multi-linear, its extrema occur when the $\xhat_i$
are from the set $\{a,b\}$ (we call such a setting of the $\xhat_i$
extremal). Note that the $\xhat_i$ are positive and $\xhat_5$ occurs only
in terms with negative sign. Thus a minimum occurs for $\xhat_5=b$.
The only extremal settings of the $\xhat_i$ which have
$\xhat_1\xhat_3+\xhat_2\xhat_4>\xhat_1\xhat_2+\xhat_3\xhat_4$ are $\xhat_1=\xhat_3=b,\xhat_2=\xhat_4=a$ and
$\xhat_1=\xhat_3=a,\xhat_2=\xhat_4=b$. For the other extremal settings (\ref{ecf})
is positive, since $b<1/2$. For $\xhat_1=\xhat_3=b,\xhat_2=\xhat_4=a$ the value of
(\ref{ecf}) is $b(a-(a^2+b^2))$.
\EPRF

\begin{remark}
In contrast to the above lemma, it is known that there is no
linear invariant for the CFN model.  This implies that there is also
no linear invariant for the restricted CFN model considered above,
since such an invariant would then extend to the general model.
This shows that the notion of linear test is more useful in some settings
than linear invariants.
\end{remark}

\section{Non-identifiability in K3}
\label{sec:K3ambiguity} \label{sec:K3}

In this section we prove there exists a non-identifiable mixture distribution in the K3 model.
Our result holds  even when the rate
matrix is the same for all edges in the
tree (the edges differ only by their associated time), i.e.,
the common rate matrix framework.
Morevoer, we will show that for most
rate matrices $R$ in the K3 model there exists a non-identifiable
mixture in which all transition matrices are generated from $R$.

The Kimura's 3-parameter model (K3) has $\Omega_{{\rm K2}}=\{0,1,2,3\}$ and
its semigroup ${\cal M}_{\rm K3}$ consists of matrices of the
following form (which we have centered around their stationary distribution):
$$
M_{\rm K3}(\xhat,\yhat,\zhat)=\frac{1}{4}E
+\left(
\begin{array}{cccc}
\xhat+\yhat+\zhat & -\xhat & -\yhat & -\zhat \\
-\xhat & \xhat+\yhat+\zhat & -\zhat & -\yhat \\
-\yhat & -\zhat & \xhat+\yhat+\zhat & -\xhat \\
-\zhat & -\yhat & -\xhat & \xhat+\yhat+\zhat
\end{array}
\right), $$ with $\xhat\leq \yhat\leq \zhat<1/4$, $\xhat+\yhat>0$, and $(\xhat+\yhat)>
2(\xhat+\zhat)(\yhat+\zhat)$.
Note that
$M_{\rm K3}(\xhat,\yhat,\yhat)=M_{\rm K2}(\xhat,\yhat)$, i.\,e., the K2 model is a
special case of the K3 model.

The group of symmetries is $G_{{\rm K3}}=\<(01)(23),(02)(13)\>$
(which is again the Klein group). There are $64$ orbits in
$\Omega^4$ under the action of $G_{\rm K3}$ (each orbit has a
representative in which $0$ appears first). The action of $K$
further decreases the number of orbits to $28$. The following
orbits are fixed by $(1\,4)$:
$$
\begin{small}
\begin{array}{l}0000\end{array},
\begin{array}{l}0110\end{array},
\begin{array}{l}0220\end{array},
\begin{array}{l}0330\end{array},
\begin{array}{l}0331\\0320\\0230\\0221\end{array},
\begin{array}{l}0332\\0310\\0130\\0112\end{array},
\begin{array}{l}0223\\0210\\0120\\0113\end{array},
\begin{array}{l}0111\\0100\\0010\\0001\end{array},
\begin{array}{l}0222\\0200\\0020\\0002\end{array},
\begin{array}{l}0333\\0300\\0030\\0003\end{array}
\end{small}
$$
The following orbits switch as indicated under the action of
$(1\,4)$:
$$
\begin{small}
\begin{array}{l}0322\\0311\\0021\\0012\end{array}
\leftrightarrow
\begin{array}{l}0232\\0201\\0131\\0102\end{array},
\begin{array}{l}0233\\0211\\0031\\0013\end{array}
\leftrightarrow
\begin{array}{l}0323\\0301\\0121\\0103\end{array},
\begin{array}{l}0133\\0122\\0032\\0023\end{array}
\leftrightarrow
\begin{array}{l}0313\\0302\\0212\\0203\end{array}
\end{small}
$$
and
$
\begin{small}
0011\leftrightarrow
0101,
0022\leftrightarrow
0202,
0033\leftrightarrow
0303,
0123\leftrightarrow
0213,
0132
\leftrightarrow
0312,
0231
\leftrightarrow
0321
\end{small}.
$

\subsection{No Linear Test for K3}
By Lemma~\ref{lem:tes} any test is a linear combination of
\begin{eqnarray*}
t_1&=&\mu(0011)-\mu(0101),\\
t_2 & = & \mu(0322)+\mu(0311)-\mu(0232)-\mu(0201)-
\\ & & \ \ \ \
\mu(0131)-\mu(0102)+\mu(0021)+\mu(0012),\\
t_3&=&\mu(0233)+\mu(0211)-\mu(0323)-\mu(0301)
\\ & & \ \ \ \
+\mu(0031)+\mu(0013)-\mu(0121)-\mu(0103),\\
 t_4&=&\mu(0022)-\mu(0202),\\
t_5&=&\mu(0133)+\mu(0122)+\mu(0032)+\mu(0023)
\\ & & \ \ \ \
-\mu(0313)-\mu(0302)-\mu(0212)-\mu(0203),\\
  t_6&=&\mu(0033)-\mu(0303),\\
t_7&=&\mu(0123)-\mu(0213),\\
 t_8&=&\mu(0132)-\mu(0312),\\ t_9&=&
 \mu(0231)-\mu(0321).
\end{eqnarray*}

We first present a non-constructive proof of non-identifiability by
proving that there does not exist a linear test, and then
 Theorem \ref{thm:dual} implies there exists a non-identifiable mixture.
We then prove the stronger result where the rate matrix is fixed.

\BLEM\label{lem:K3-ambiguity}
There does not exist a linear test for the K3 model.
\ELEM

\BCOR
There exists a non-identifiable mixture in the K3 model.
\ECOR

\begin{proof}[Proof of Lemma \ref{lem:K3-ambiguity}]
 Suppose that $t=\alpha_1 t_1+\dots+\alpha_9 t_9$ is a test.
 Let $\what_i=(\xhat_i,\yhat_i,\zhat_i)$, $1\le i\le 5$.  For $1\le i\le 4$,
 $\what_i$ denotes the centered parameters for the edge
 incident to leaf $i$, and $\what_5$ are the centered parameters for the internal edge.

In the definitions of $\mu_1$ and $\mu_2$ below we will set
$\what_2=\what_4=\what_5=(0,0,0)$. This ensures that
in labelings with non-zero probability, leaves $v_2, v_4$ and both internal
vertices all have the same label.
Moreover, by
observation~\ref{obs:zero}, $\Phi(\mu_i)=0$.

Let $\mu_1$
be generated from $T_2$ with $\what_1=(1/4,\yhat_1,\zhat_1)$, and
$\what_3=(1/4,0,0)$. In labelings with non-zero probability, the
labels of $v_2$ and $v_3$ have to both be in $\{0,1\}$ or both in
$\{2,3\}$. The only labels in $t_1,\dots,t_9$ with this property
are $0101,0232,0323$. Thus,
\begin{equation}\label{eq:q2c}
t^T\mu_1=-\alpha_1\mu(0101)-\alpha_2\mu_1(0232)-\alpha_3\mu_1(0323)=
-\frac{1}{16}(\alpha_1/4+\alpha_2 \yhat_1+\alpha_3 \zhat_1)=0.
\end{equation}
Any $\yhat,\zhat\in[1/8-\varepsilon,1/8+\varepsilon]$ with $y'\geq z'$ gives
a valid matrix in the K3 model. For (\ref{eq:q2c}) to be always
zero we must have $\alpha_1=\alpha_2=\alpha_3=0$ (since $1,\yhat_1,\zhat_1$
are linearly independent polynomials).

Let $\mu_2$ be generated from $T_2$ with $\what_1=(1/4,\yhat_1,\zhat_1)$, and
$\what_3=(1/4,\yhat,\zhat)$. The only labels in $t_4,\dots,t_9$ with $v_2$ and
$v_4$ having the same label are
\[ 0202,0313,0212,0303\]  (we ignore
the labels in $t_1,t_2,t_3$ because
$\alpha_1=\alpha_2=\alpha_3=0$). Thus
\begin{equation}\label{eq:q2b}
\begin{split}
t^T\mu_1=-\alpha_4\mu_2(0202)-\alpha_5\mu_2(0313)-
\alpha_5\mu_2(0212)-\alpha_6\mu_2(0303)  \\
= -\frac{1}{4}(\alpha_4 (\yhat_1)^2 + 2\alpha_5 \yhat_1\zhat_1 + \alpha_6 (\zhat_1)^2)=0.
\end{split}
\end{equation}
Polynomials $(\yhat_1)^2,\yhat_1\zhat_1,(\zhat_1)^2$ are linearly independent and hence
$\alpha_4=\alpha_5=\alpha_6=0$.

Thus $t$ is a linear combination of $t_7,t_8,t_9$ and hence has at
most $6$ terms. A test for K3 must be a test for K2, but the
unique test for K2 has $8$ terms. Thus $\Phi$ cannot be a test for
K3.

\end{proof}

\subsection{Non-identifiable mixtures in the K3 model with a fixed rate matrix}

We will now consider rate matrices for the model, as opposed
to the transition probability matrices.
The rate matrix for the K3 model is
\begin{equation}\label{krt}
R=R(\alpha,\beta,\gamma)=\left(
\begin{array}{cccc}
-\alpha-\beta-\gamma & \alpha & \beta & \gamma \\
\alpha & -\alpha-\beta-\gamma & \gamma & \beta \\
\beta & \gamma & -\alpha-\beta-\gamma & \alpha \\
\gamma & \beta & \alpha & -\alpha-\beta-\gamma
\end{array}
\right),
\end{equation}
where the rates usually satisfy $\alpha\geq\beta\geq\gamma> 0$. For our examples
we will also assume that $\alpha,\beta,\gamma\in [0,1]$. Since the examples are
negative they work immediately for the above weaker constraint.

Recall, the K2 model is a submodel of the K3 model: the rate matrices of
the K2 model are precisely the rate matrices of K3 model with
$\beta=\gamma$. By Lemma~\ref{lk2} there exists a test in the K2
model and hence there are no non-identifiable mixtures. We will show that
the existence of a test in the K2 model is a rather singular
event: for almost all rate matrices in the K3 model there exist
non-identifiable mixtures and hence no test.

We show the following result.

\BLEM\label{rara} Let $\alpha,\beta,\gamma$ be chosen independently from
the uniform distribution on $[0,1]$. With probability $1$ (over the choice of $\alpha,\beta,\gamma$)
there does not exist a test for the K3 model with the rate matrix
$R(\alpha,\beta,\gamma)$.
\ELEM

To prove Lemma~\ref{rara} we need the following technical concept.
A {\em generalized polynomial} is a function of the form
\begin{equation}\label{form}
\sum_{i=1}^m a_i(u_1,\dots,u_n) e^{b_i(u_1,\dots,u_n)},
\end{equation}
where the $a_i$ are non-zero polynomials and the $b_i$ are
distinct linear polynomials. Note that the set of generalized
polynomials is closed under addition, multiplication, and taking
derivatives. Thus, for example, the Wronskian of a set of
generalized polynomials (with respect to one of the variables, say $u_1$)
is a generalized polynomial.

For $n=1$ we have the following bound on the number of roots of a
generalized polynomial (see \cite{PS76-2}, part V, problem 75):

\BLEM\label{lem:bnd} Let $G=\sum_{i=1}^m a_i(u) e^{b_i(u)}$ be a univariate generalized
polynomial. Assume $m>0$. Then $G$ has at most
$$(m-1)+\sum_{i=1}^m{\rm deg}\, a_i(u)$$
real roots, where ${\rm deg}\,a_i(u)$ is the degree of the polynomial $a_i(u)$.
\ELEM

\BCOR\label{comes} Let $G(u_1,\dots,u_n)=\sum_{i=1}^m
a_i(u_1,\dots,u_n) e^{b_i(u_1,\dots,u_n)}$ be a generalized
polynomial. Assume $m>0$ (i.\,e., that $G$ is not the zero
polynomial). Let $r_1,\dots,r_n$ be picked in independently and
randomly from uniform distribution on $[0,1]$. Then with
probability $1$ (over the choice of $r_1,\dots, r_n$)
we have $G(r_1,\dots,r_n)\neq 0$. \ECOR

\begin{proof}[Proof of Corollary~\ref{comes}]
We will proceed by induction on $n$. The base case $n=1$ follows
from Lemma~\ref{lem:bnd} since the probability of a finite set is zero.

For the induction step consider the polynomial $a_1$. Since $a_1$ is a
non-zero polynomial there are only
finitely many choices of $c\in [0,1]$ for which $a_1(c,u_2,\dots,u_n)$
is the zero polynomial. Thus with
probability $1$ over the choice of $u_1=c$ we have that
$a_1(c,u_2,\dots,u_n)$ is a non-zero polynomial. Now,
by the induction hypothesis, with probability $1$ over the choice of
$u_2,\dots,u_n$ we have that
$G(c,u_2,\dots,u_n)\neq 0$. Hence with probability $1$ over the choice
of $u_1,\dots,u_n$ we have
that $G(u_1,\dots,u_n)\neq 0$.
\end{proof}


\begin{proof}[Proof of Lemma~\ref{rara}]
Any test has to be a linear combination of $t_1,\dots,t_9$
defined in Section~\ref{sec:K3ambiguity}. Suppose that $t=\sigma_1 t_1+\dots
\sigma_9 t_9$ is a test. We will use Observation~\ref{obs:zero} to
show that $\sigma_1=\dots=\sigma_9=0$ and hence $t$ cannot be a
test. Let $\sigma=(\sigma_1,\dots,\sigma_9)$.

The transition matrix of the process with rate matrix $R$, at time
$s$ is $T(s)=\exp(sR)$. We have (see, e.g., \cite{F:book})
\begin{equation}\label{dek}
T(s)=M_{K3}=
\frac{1}{4}E+
\left(
\begin{array}{cccc}
A+B+C & -A & -B & -C \\
-A & A+B+C & -C & -B \\
-B & -C & A+B+C & -A \\
-C & -B & -A & A+B+C
\end{array}
\right),
\end{equation}
where
\begin{eqnarray*}
A=(1+e^{-2s(\alpha+\beta)}+e^{-2s(\alpha+\gamma)}-e^{-2s(\beta+\gamma)})/4,\\
B=(1+e^{-2s(\alpha+\beta)}-e^{-2s(\alpha+\gamma)}+e^{-2s(\beta+\gamma)})/4,\\
C=(1-e^{-2s(\alpha+\beta)}+e^{-2s(\alpha+\gamma)}+e^{-2s(\beta+\gamma)})/4.
\end{eqnarray*}
We have changed notation from our earlier definition of $M_{K3}$ by exchanging
$A,B,C$ for $\xhat,\yhat,\zhat$ to indicate these as functions of $\alpha,\beta$ and
$\gamma$.

Let $\mu_x$ be generated from $T_3$ with edge weights
$T(x),T(2x),T(3x),T(4x),T(0)$. The internal edge has length $0$
(i.\,e., it is labeled by the identity matrix), and hence, by
Observation~\ref{obs:zero}, we must have $t^T\mu_x=0$ for all $x\geq 0$.

Let $f_i(x)=t_i^T\mu_x$ for $i=1,\dots,9$. Let $W(x)$ be the
Wronskian matrix of $f_1(x),\dots,f_9(x)$ with respect to $x$. The entries
of $W(x)$ are generalized polynomials in variables
$\alpha,\beta,\gamma$, and $x$. Thus the Wronskian (i.\,e., $\det
W(x)$) is a generalized polynomial in variables
$\alpha,\beta,\gamma$, and $x$.

Now we show that for a particular choice of $\alpha,\beta,\gamma$
and $x$ we have $\det W(x)\neq 0$. Let $\alpha=\pi i$, $\beta=2\pi
i$, $\gamma=-2\pi i$, and $x=1$. Of course complex rates are not
valid in the K3 model, we use them only to establish that $W(x)$
is a non-zero generalized polynomial. Tedious computation (best
performed using a computer algebra system) yields that the
Wronskian matrix is the following.
The first 7 columns are
\begin{tiny}
$$
\left(
\begin{matrix}
8& -16& -16& 8& 16& 8& 0
\\
40i\pi& -240i\pi& 80i\pi& 56i\pi& 80i\pi& 24i\pi& 48i\pi
\\
-1052\pi^2& 3736\pi^2& 856\pi^2& -780\pi^2& -824\pi^2& -300\pi^2& 112\pi^2
\\
-12244i\pi^3& 52920i\pi^3& -4520i\pi^3& -13988i\pi^3& -2600i\pi^3& 732i\pi^3& -4224i\pi^3
\\
206468\pi^4& -923656\pi^4& -134536\pi^4& 187188\pi^4& 46856\pi^4& 9108\pi^4& 1520\pi^4
\\
3199060i\pi^5& -12454200i\pi^5& -280600i\pi^5& 3574316i\pi^5& 103400i\pi^5& -354276i\pi^5& 576168i\pi^5
\\
-46789172\pi^6& 230396776\pi^6& 25447336\pi^6& -47807340\pi^6& -2740904\pi^6& 1992900\pi^6& -2781848\pi^6
\\
-826613044i\pi^7& 3116686680i\pi^7& 184720120i\pi^7& -908923988i\pi^7& -4488200i\pi^7& 82589532i\pi^7& -98067024i\pi^7
\\
11742908228\pi^8& -58371817096\pi^8& -5047384456\pi^8& 12445876068\pi^8& 163759496\pi^8& -776675292\pi^8& 840993440\pi^8
\end{matrix}
\right.
$$
\end{tiny}
The last 2 columns are
\begin{tiny}$$
\left.
\begin{matrix}
0& 0 \\
-48i\pi& -96i\pi \\
912\pi^2& 800\pi^2\\
14880i\pi^3& 17568i\pi^3\\
-212400\pi^4& -198560\pi^4\\
-3692808i\pi^5&   -4100016i\pi^5\\
49984152\pi^6&  51191600\pi^6\\
919913280i\pi^7&  1003352208i\pi^7\\
-12613821600\pi^8&  -13323533120\pi^8
\end{matrix}.
\right)
$$
\end{tiny}
The determinant of $W(x)$ is
$33920150890618370745095852723798016000000\pi^{36}$ which is
non-zero. Thus $\det W(x)$ is a non-zero generalized polynomial
and hence for random $\alpha,\beta,\gamma$ and $x$ we have that
$\det W(x)$ is non-zero with probability $1$.

Assume now that $\det W(x)\neq 0$. Let $w=W(x)\sigma$. The first
entry $w_1$ of $w$ is given by $w_1=t^T\mu_x$. Since $t$ is a test
we have $w_1=0$. The second entry $w_2$ of $w$ is given by
$w_2=(\partial/\partial x)t^T\mu_x$. Again we must have $w_2=0$,
since $t$ is a test. Similarly we show $w_3=\dots=w_9=0$. Note
that $W(x)$ is a regular matrix and hence $\sigma$ must be the
zero vector. Thus $t$ is not a test. Since $\det W(x)\neq 0$
happens with probability $1$ we have that there exists no test
with probability $1$.
\end{proof}

\section{Maximum Likelihood Results}
\label{sec:maxlik}

Here we prove Theorem \ref{thm:MLEbad}.

\subsection{Terminology}

In the CFN and JC model there is a single parameter defining the transition matrix
for each edge, namely a parameter $x_e$ where $0<x_e<1/|\Omega|$.
In K2, there are 2 parameters and in K3 there are 3 parameters.
We use the term weights $\vec{w}_e$ to denote the setting of the parameters defining the
transition matrix for the edge $e$.  And then we let $\vec{w}=\left(\vec{w}_e\right)_{e\in T}$
denote the set of vectors defining the transition matrices for edges of tree $T$.

We use the term zero weight edge to denote the setting of the weights so that the
transition matrix is the identity matrix $I$.  Thus, in the CFN and JC models, this
corresponds to setting $x_e=0$.  We refer to a non-zero weight edge as a
setting of the parameters so that all entries of the transition matrix are positive.
Hence, in the CFN and JC models this corresponds to $x_e>0$.

Note, ${\cal L_T}(\mu)$ is maximized over the set of weights $\vec{w}$.
Hence our consideration of weights $\vec{w}$.  There are typically constraints
on $\vec{w}$ in order to define valid transition matrices.
For example,
in the CFN model we have $\vec{w}_e=x_e$ where
$0<x_e< 1/2$, and similarly $0<x_e<1/4$ in the JC model.
In the K2 model we have $\vec{w}_e=(x,y)$ where $x\geq y>0, x+y<1/2$.
Finally in the K3 model we have further constraints as detailed in Section \ref{sec:K3}.

\subsection{Technical Tools}

Our proof begins starts from the following observation.
Consider the CFN model.
Note that for $x=0$ we
have $\vec{P_1}=\vec{P_2}'$ and hence $\mu_0=\mu_{T_3,\vec{v}}$
where $\vec{v}=(1/2,1/2,1/2,1/2,0)$ are the weights for the CFN model,
i.e., $\mu_0$ is generated by a pure distribution from $T_3$. In
fact it can be generated by a pure distribution from any
leaf-labeled tree on $4$ leaves since the internal edges have zero length.

\BOBS\label{obs:achievable}
Consider a tree $T$ on $n$ leaves and weights $w$ where all internal edges have
zero weight.  For all trees $S\neq T$ on $n$ leaves, there is a unique weight $\vec{v}$
 such that $\mu_{S,\vec{v}}=\mu_{T,\vec{w}}$.
\EOBS

\BPRF
Let $\vec{v}$ have the same weight as $\vec{w}$ for all terminal edges, and zero weight for
all internal edges.   Note,
we then have $\mu_{S,\vec{v}}=\mu_{T,\vec{w}}$ and it remains to prove uniqueness of
$\vec{v}$.

Let $\vec{u}$ be a set of weights where $\mu_{S,\vec{u}}=\mu_{S,\vec{w}}$.  Let $S'$ be obtained
from $S$ by contracting all the edges of zero weight in $\vec{u}$, and let $\vec{u'}$ be the
resulting set of weights for the remaining edges. The tree $S'$ has all
internal vertices of degree $\geq 3$ and internal edges have non-zero weight.

It follows from the work of Buneman \cite{Buneman} that
$S',\vec{u'}$ is unique among trees with non-zero internal edge weights and without vertices
of degree two.   If $\vec{u}=\vec{v}$, then $S'$ is a star, and hence every $\vec{u}$ must contract
to a star.  This is only possible if $\vec{u}$ assigns zero weight to all internal edges.
Therefore, $\vec{v}$ is the unique weight.
\end{proof}

For $w$ and $w'$ defined for the CFN model
in Theorem \ref{thm:MLEbad},  for any 4-leaf tree $S$,  the maximum of ${\cal L}_S(\mu_0)$ is achieved on
$v=(1/4,1/4,1/4,1/4,0)$. For any $v'\neq v$ the distribution
$\mu_{S,v'}$ is different from $\mu_0$ and hence ${\cal
L}_{S,v'}(\mu_0)<{\cal L}_{S,v}(\mu_0)=\mu_0^T\ln\mu_0$. Intuitively, for small
$x$ the maximum of ${\cal L}_S(\mu_x)$ should be realized on a
$v''$ which is near $v$.  Now we formalize this argument in the following lemma.
Then in Lemma \ref{lem:mai} we will use the
Hessian and Jacobian of the expected log-likelihood functions to bound
${\cal L}_S(\mu_x)$ in terms of ${\cal L}_S(\mu_0)$.

 \BLEM\label{obs:epsdel}
Let $\mu$ be a probability distribution on $\Omega^n$ such that
every element has non-zero probability. Let $S$ be a leaf-labeled
binary tree on $n$ leaves. Suppose that there exists a unique
$v$ in the closure of the model such that $\mu_{S,v}=\mu$.
 Then for every $\delta>0$ there exists $\eps>0$ such
that for any $\mu'$ with $||\mu'-\mu||\leq\eps$ the global optima
of ${\cal L}_S(\mu')$ are attained on $v'$ for which
$||v'-v||\leq\delta$.
 \ELEM

\begin{remark}
In our application of the above lemma we have $\mu=\mu_0$.
Consider a tree $S$ and its unique weight $v$ where $\mu_{S,v}=\mu_0$.
Note, the requirement that every element in $\{0,1\}^n$ has
non-zero probability, is satisfied if the terminal edges have non-zero weights.
In contrast, the internal edges have zero weight so that Observation \ref{obs:achievable}
applies, and in some sense the tree is achievable on every topology.
\end{remark}

\BPRF
We will prove the lemma by contradiction.
Roughly speaking, we now
suppose that there exists $\mu'$ close to $\mu$ where
the maximum of $\mu'$ is achieved far from the maximum of $\mu$.
Formally, suppose that there exists $\delta>0$ and sequences
$\mu'_i$ and $v'_i$ such that $\lim_{i\ra\infty}
||\mu-\mu'_i||=0$ and $||v'_i-v||>\delta$ where $v'_i$ is a
weight for which the optimum of ${\cal L}_S(\mu'_i)$ is attained.
By the optimality of the $v_i'$ we have
\begin{equation}\label{e:1}
{\cal L}_{S,v_i'}(\mu_i')\geq{\cal L}_{S,v}(\mu_i').
\end{equation}
We assumed $\mu_{S,v}=\mu$ and hence the entries of
$\ln\mu_{S,v}$ are finite since we assumed $\mu$ has
positive entries. Thus
\begin{equation}\label{e:2}
\lim_{i\ra\infty} {\cal
L}_{S,v}(\mu'_i)= \lim_{i\ra\infty} ({\mu'}_i^T \ln\mu_{S,v}) = \mu^T
\ln\mu_{S,v}={\cal L}_{S,v}(\mu).
\end{equation}
Take a subsequence of the $v'_i$ which converges to some $v'$.  Note,
$v'$ is in the closure of the model.

Let
$\eps$ be the smallest entry of $\mu$.
For all sufficiently large
$i$, $\mu'_i$ has all entries $\geq\eps/2$.

Hence,
\begin{equation}
\label{eq:zzzz}  {\cal L}_{S,v'_i}(\mu_i') =
   \sum_{z\in\Omega^n}\mu'_i(z)\ln(\mu_{S,v'_i}(z))
   \leq
    \frac{\eps}{2}\min_{z\in\Omega^n}\ln(\mu_{S,v'_i}(z))
   \end{equation}

Because of \eqref{e:2}, for all sufficiently large
$i$,
${\cal L}_{S,v}(\mu_i')\geq 2{\cal L}_{S,v}(\mu)$
(recall that the log-likelihoods are negative).
Combining with \eqref{eq:zzzz}, we have
that the entries of $\ln\mu_{S,v'_i}$ are bounded from
below by  ${\cal L}_{S,v}(\mu)/4\eps$.  Thus, both $\mu'_i$ and $\ln\mu_{S,v'_i}$
are bounded.  For bounded and convergent sequences,
\[
\lim_{n\rightarrow\infty} a_nb_n =
\lim_{n\rightarrow\infty} a_n
\lim_{n\rightarrow\infty} b_n
\]
Therefore,
\begin{equation}\label{e:3}
\lim_{i\ra\infty} {\cal L}_{S,v'_i}(\mu_i')=
(\lim_{i\ra\infty}{\mu'}^T_i)(\lim_{i\ra\infty} \ln\mu_{S,v'_i}) =
\mu^T \ln\mu_{S,v'}={\cal L}_{S,v'}(\mu).
\end{equation}
From (\ref{e:1}), (\ref{e:2}), and (\ref{e:3}) we have
${\cal L}_{S,v'}(\mu)\geq{\cal L}_{S,v}(\mu)$. Since $v'\neq v$
we get a contradiction with the uniqueness of $v$.
\EPRF

We now bound the difference of  ${\cal L}_S(\mu')$
and ${\cal L}_S(\mu)$ when the previous lemma applies.
This will then imply that for $x$ sufficiently small,
${\cal L}_S(\mu_x)$ is close to ${\cal L}_S(\mu_0)$.

Here is the formal statement of the lemma.

\begin{lemma} \label{lem:mai}
\label{lem:SV} Let $\mu\in{\mathbb R}^{|\Omega|^n}$ be a probability distribution on $\Omega^n$ such that
every element has non-zero probability. Let $S$ be a leaf-labeled binary tree on $n$ vertices. Suppose that
there exists $\vec{v}$ in the closure of the model such that $\mu_{S,\vec{v}}=\mu$ and that
$\vec{v}$ is the unique such weight.
Let $\Delta\mu_x\in{\mathbb R}^{|\Omega|^n}$ be such that $\Delta\mu_x^T 1=0$, and $x\mapsto\mu_x$ is
continuous around $0$ in the following sense: $\Delta\mu_x\ra\Delta\mu_0$ as $x\ra 0$.

Let $g(\vec{w})={\cal L}_{S,\vec{w}}(\mu)$, and $h_x(\vec{w})=(\Delta\mu_x)^T\ln\mu_{S,\vec{w}}$.
Let $H$ be the Hessian of $g$ at $\vec{v}$ and $J_x$ be the Jacobian
of $h_x$ at $\vec{v}$. Assume that $H$ has full rank. Then
\begin{equation}\label{e:es}
{\cal L}_{S}(\mu+x\Delta\mu_x)\leq \mu^T\ln\mu + xh_x(\vec{v})-
\frac{x^2}{2}J_0H^{-1}J_0^T+o(x^2).
\end{equation}
Moreover, if $(H^{-1}J^T)_i\leq 0$ for all $i$ such that $\vec{v}_i=0$ then
 the inequality in (\ref{e:es}) can be replaced by equality.
\end{lemma}

\begin{remark}
\label{rem:nonzero}
When $(H^{-1}J^T)_i< 0 $ for all $i$ such that $v_i=0$ then
the likelihood is maximized at non-trivial branch lengths.  In particular,
for the CFN model, the branch lengths are in the interval $(0,1/2)$,
that is, there are no branches of length $0$ or $1/2$.
Similarly for the Jukes-Cantor model the lengths are in $(0,1/4)$.
\end{remark}

 \BPRF
 For notational convenience,
let $f(\vec{w})=\ln\mu_{S,\vec{w}}$.  Thus, $g(\vec{w}) = \mu^T f(\vec{w})$. Note
that
\begin{equation}\label{e:aaa1}
f(\vec{v})=\ln\mu.
\end{equation}
 The function $f$ maps
assignments of weights for the $2n-3$ edges of $S$ to the logarithm of the distribution induced on the
leaves. Hence, the domain of $f$ is the closure of the model, which is a subspace of $\real^{d(2n-3)}$, where
$d$ is the dimension of the parameterization of the model, e.g., in the CFN and JC models $d=1$ and in the K2
model $d=2$.  We denote the closure of the model as $\Lambda$.  Note, the range of $f$ is
$[-\infty,0]^{|\Omega|^n}$.

 Let $J_f=(\partial f_i/\partial w_j)$ be the Jacobian of $f$ and
$H_f=(\partial f_i/\partial w_j\partial w_k)$ be the Hessian of $f$ ($H_f$ is a rank $3$ tensor).

If $\vec{v}$ is in the interior of $\Lambda$, then since $\vec{w}$ optimizes
${\cal L}_{S,\vec{v}}(\mu)$ we have
\begin{equation}\label{e:j}
\mu^T J_f(\vec{v})=0.
\end{equation}
We will now argue that equality (\ref{e:j}) remains true even
when $\vec{v}$ is on the boundary
of $\Lambda$. Function $\mu_{S,\vec{v}}$ is a polynomial function
of the coordinates of $\vec{v}$. The coordinates of $\mu_{S,\vec{v}}$ sum to the constant $1$
function. We assumed that $\mu=\mu_{S,\vec{v}}$ is strictly positive and
hence for all $\vec{w}$ in a small neighborhood of $\vec{v}$ we have that $\mu_{S,\vec{w}}$
is still a distribution (note that $\vec{w}$ can be outside of
$\Lambda$).  Suppose that $\mu^TJ_f(\vec{v})\neq 0$. Then there exists  $\vec{\Delta v}$
such that $\mu^T\ln\mu_{S,\vec{v}+\vec{\Delta v}}>\mu^T\ln\mu_{S,\vec{v}}=\mu^T\ln\mu$.
This contradicts the fact that
the maximum (over all distributions $\nu)$ of $\mu^T\ln(\nu)$
is achieved at $\nu=\mu$.
 Thus (\ref{e:j}) holds.

If we perturb $\mu$ by $x\Delta\mu_x$ and $\vec{v}$ by $\vec{\Delta v}$ the
likelihood changes as follows:
\begin{eqnarray}
\nonumber
\lefteqn{   {\cal L}_{T,\vec{v}+\vec{\Delta v}}(\mu+x\Delta\mu_x) } \\
&=&   (\mu+x\Delta\mu_x)^T f(v+\Delta v)
\nonumber
\\
&=&
(\mu+ x\Delta\mu_x)^T\Big(f(\vec{v})+\left(J_f(\vec{v})\right)(\vec{\Delta v})
+\frac{1}{2}(\vec{\Delta v})^T \left(H_f(\vec{v})\right) (\vec{\Delta v})+
O(||\vec{\Delta v}||^3)\Big)
\nonumber
\\
&=&
\mu^T\ln\mu+\mu^T\left(\left(J_f(\vec{v})\right)(\vec{\Delta v})+\frac{1}{2}(\vec{\Delta v})^T H_f(\vec{v})(
\vec{\Delta v})\right)
\nonumber
\\ && +
x (\Delta\mu_x)^T\left(
f(\vec{v})+J_f(\vec{v})(\vec{\Delta v})
\right)+O\left(||\vec{\Delta v}||^3+x||\vec{\Delta v}||^2\right)
\nonumber
\\
&=&
\mu^T\ln\mu+\mu^T\Big(\frac{1}{2}(\Delta v)^T H_f(\vec{v})(
\vec{\Delta v})\Big)+ x(\Delta\mu_x)^T\Big(f(\vec{v})+J_f(\vec{v})(\vec{\Delta v})
\Big)
\nonumber
\\  && +O\left(||\vec{\Delta v}||^3+x||\vec{\Delta v}||^2\right),
\label{e:22}
\end{eqnarray}
where in the third step we used \eqref{e:aaa1} and in the last step we used (\ref{e:j}). In terms of $H,J$
defined in the statement of the theorem we have
\begin{multline}
{\cal L}_{T,\vec{v}+\vec{\Delta v}}(\mu+x\Delta\mu_x)=
\mu^T\ln\mu+x h_x(\vec{v})+\frac{1}{2}(\vec{\Delta v})^T H(\vec{\Delta v})
+x J_x(\vec{\Delta v}) \\ +O\left(||\vec{\Delta v}||^3+x||\vec{\Delta v}||^2\right).
\label{eq:27}
 \end{multline}

 We will now prove that $H$ is negative definite.  First note that
 we assumed that $H$ is full-rank, thus all of its eigenvalues are non-zero.
 Moreover, $g(\vec{w})={\cal L}_{S,\vec{w}}(\mu)$
 is maximized for $\vec{w}=\vec{v}$.
Plugging $x=0$ into \eqref{eq:27}, we obtain
\begin{equation}
g(\vec{v} + \vec{\Delta v}) =
\mu^T\ln\mu+\frac{1}{2}(\vec{\Delta v})^T H(
\vec{\Delta v})+ O\left(||\vec{\Delta v}||^3\right)
\label{eq:www}
\end{equation}
Hence, if $H$ was not negative definite, then it would have at least one
positive eigenvalue.  Let $\vec{z}$ denote the corresponding eigenvector.
 Let $\vec{\Delta v}$ be a sufficiently small multiple of $\vec{z}$.
 By \eqref{eq:www} we have $g(\vec{v} + \vec{\Delta v})>g(\vec{v})$ which contradicts the
earlier claim that $\vec{w}=\vec{v}$ maximizes $g(\vec{w})$.
Hence, all of the eigenvalues of $H$ are negative, i.e., it is negative definite.
Let $\lambda_{\rm max}$ be the largest
eigenvalue of $H$, i.e., closest to zero. We will use that $\lambda_{\rm max}<0$.
Note,
\begin{equation}\label{eq:eq}
(\vec{\Delta v})^T H(\vec{\Delta v}) \leq  \lambda_{\max}||\vec{\Delta v}||^2
\end{equation}
Set $\vec{\Delta v}$ so that
${\cal L}_{\vec{v}+\vec{\Delta v}}(\mu+x\vec{\Delta}\mu_x)$ is maximized.
($\vec{\Delta v}$ is a function of $x$.)
Now we will prove that $||\vec{\Delta v}|| = O(x)$.
For any $\delta>0$, by
Lemma~\ref{obs:epsdel}, for all sufficiently small $x$, we have
$||\vec{\Delta v}||\leq\delta$.
By \eqref{eq:27} and \eqref{eq:eq} we have
\begin{multline*}
{\cal L}_{S,\vec{v}+\vec{\Delta v}}(\mu+x\Delta\mu_x) \leq
\mu^T\ln\mu+x h_x(\vec{v})+\lambda_{\max} ||\vec{\Delta v}||^2
+x J_x(\vec{\Delta v}) \\
+O(||\vec{\Delta v}||^3+x||\vec{\Delta v}||^2).
 \end{multline*}

Assume $x=o(||\vec{\Delta v}||)$ (i.e., assume that $||\vec{\Delta v}||$
goes to zero more slowly than $x$).
Then, we have
\[
{\cal L}_{S,\vec{v}+\vec{\Delta v}}(\mu+x\Delta\mu) \leq
\mu^T\ln\mu+x h_x(\vec{v})+\lambda_{\max} ||\vec{\Delta v}||^2
+o(||\vec{\Delta v}||^2).
 \]
On the other hand, we have
\[
{\cal L}_{S,\vec{v}}(\mu+x\Delta\mu) = (\mu+x\Delta\mu)^T\ln\mu_{S,\vec{v}} =
\mu^T\ln\mu+x h_x(\vec{v}).
\]
Since $\lambda_{\max} ||\vec{\Delta v}||^2$ is negative then for sufficiently small
$||\vec{\Delta v}||$ (recall we can choose any $\delta>0$ where $||\vec{\Delta v}||\leq\delta$),
we have
\[
{\cal L}_{\vec{v}+\vec{\Delta v}}(\mu+x\Delta\mu) \leq
{\cal L}_{\vec{v}}(\mu+x\Delta\mu).
\]

Thus we may restrict ourselves to $\vec{\Delta v}$
such that $||\vec{\Delta v}||=O(x)$. Hence
 \begin{multline}\label{e:max}
{\cal L}_{S}(\mu+x\Delta\mu_x)\leq
\mu^T\ln\mu+x h_x(\vec{v})  \\
+\max_{\vec{\Delta w}}\left(\frac{1}{2}(\vec{\Delta w})^T H(\vec{\Delta w})
     +x J_x(\vec{\Delta w})\right)
+O(x^3).
 \end{multline}

 The maximum of
\begin{equation}\label{e:ce}
\frac{1}{2}(\vec{\Delta w})^T H(\vec{\Delta w})+x J_x(\vec{\Delta w})
\end{equation}
occurs at $\vec{\Delta} z:=-xH^{-1}J_x^T$;
 for this $\vec{\Delta z}$ the value
of (\ref{e:ce}) is $-\frac{x^2}{2}J_x H^{-1} J_x^T$.
Therefore,
 \[
{\cal L}_{S}(\mu+x\Delta\mu)\leq
\mu^T\ln\mu+x h_x(\vec{v})  -\frac{x^2}{2}J_x H^{-1} J_x^T
+ O(x^3).
\]
From $\mu_x\ra\mu_0$, we have $
J_x H^{-1} J_x^T = (1+o(1)) J_0 H^{-1} J_0^T$, and hence
 \[
{\cal L}_{S}(\mu+x\Delta\mu)\leq
\mu^T\ln\mu+x h_x(\vec{v})  -\frac{x^2}{2}J_0 H^{-1} J_0^T
+ o(x^2).
\]
This completes the proof of the first part of the lemma.
It remains to prove the case when the inequality can be replaced by equality.

 Note that in general the inequality cannot be replaced by
equality, since $\vec{v}+x\vec{\Delta z}$ can be an invalid weight (i.\,e.,
outside of $\Lambda$) for all $x$.
 If $(\vec{\Delta v})_i\geq 0$ whenever $\vec{v}_i=0$ then $\vec{v}+x\vec{\Delta z}$ is a
valid weight vector for sufficiently small $x$ and hence
plugging directly into \eqref{eq:27} we have
\begin{equation}\label{e:ff}
\begin{split}
{\cal L}_{S}(\mu+x\Delta\mu_x)\geq \mu^T\ln\mu + xh_x(\vec{v})-
\frac{x^2}{2}J_xH^{-1}J_x^T+O(x^3)=\\
\mu^T\ln\mu + xh_x(\vec{v})-
\frac{x^2}{2}J_0H^{-1}J_0^T+o(x^2).
\end{split}
\end{equation}
\EPRF

\subsection{Proof of Theorem \ref{thm:MLEbad} in CFN: $C=1/4$}
In this section we deal with the CFN model.
We prove Part \ref{thm:CFNbad} of Theorem~\ref{thm:MLEbad}.
For simplicity, we first present the proof for the case $C=1/4$.

Let $\mu_1$ be
generated from $T_3$ with weights $\vec{P_1}=(1/4+x,1/4-x,1/4-x,1/4+x,x^2)$
and $\mu_2$ be generated from $T_3$ with weights
$\vec{P_2}=(1/4-x,1/4+x,1/4+x,1/4-x,x^2)$. Let $\mu=(\mu_1+\mu_2)/2$. Note
that $(1\,4)(2\,3)$ fixes $\mu_1$, and $\mu_2$; and $(1\,2)(3\,4)$
swaps $\mu_1$ and $\mu_2$. Hence $\mu$ is invariant under
$K=\<(1\,2)(3\,4),(1\,4)(2\,3)\>$.  This simplifies many of the following calculations.

One can verify that the Hessian is the same for all the trees:
$$H=\begin{pmatrix}
\frac{-1552}{615}  & \frac{-16}{41}  & \frac{-16}{41}  & \frac{-16}{41}  & \frac{-80}{123} \vspace{0.1cm}\\
\frac{-16}{41}  & \frac{-1552}{615}  & \frac{-16}{41}  & \frac{-16}{41}  & \frac{-80}{123} \vspace{0.1cm}\\
\frac{-16}{41}  & \frac{-16}{41}  & \frac{-1552}{615}  & \frac{-16}{41}  & \frac{-80}{123} \vspace{0.1cm}\\
\frac{-16}{41}  & \frac{-16}{41}  & \frac{-16}{41}  & \frac{-1552}{615}  & \frac{-80}{123} \vspace{0.1cm}\\
\frac{-80}{123}  & \frac{-80}{123}  & \frac{-80}{123}  & \frac{-80}{123}  & \frac{-400}{369}
\end{pmatrix}.$$
The above is straightforward to verify in any symbolic algebra system, such
as Maple.

The Jacobians differ only in their last coordinate.
For $T_3$ we have
$$J_0=\left(
\frac{1744}{615},
\frac{1744}{615},
\frac{1744}{615},
\frac{1744}{615},
\frac{-880}{369}\right),   $$
Finally,
$$-\frac{1}{2}J_0 H^{-1} J_0^T =\frac{36328}{1845} \approx 19.68997.$$

Hence, for $v=(1/4,1/4,1/4,1/4,0)$, by Lemma \ref{lem:mai},  we have
\begin{eqnarray*}
{\cal L}_{S}(\mu_x) &\leq &
 \mu^T\ln\mu + xh(v)-  \frac{x^2}{2}J_0H^{-1}J_0^T+O(x^3)
 \\
 &= &
 \mu^T\ln\mu + xh(v)
 + x^2\frac{36328}{1845}
+ O(x^3)
\end{eqnarray*}

For $T_2$ we have
$$J_=\left(
\frac{1744}{615},
\frac{1744}{615},
\frac{1744}{615},
\frac{1744}{615},
\frac{-1208}{369}\right),    $$
Then,
$$-\frac{1}{2}J_0 H^{-1} J_0^T =\frac{244862}{9225} \approx 26.54331.$$
By Lemma \ref{lem:mai},  we have
\[
{\cal L}_{S}(\mu_x)
 \leq
 \mu^T\ln\mu + xh(v)
 +  x^2\frac{244862}{9225}
+ O(x^3)
\]

For $T_1$ we have
$$J_0=\left(
\frac{1744}{615},
\frac{1744}{615},
\frac{1744}{615},
\frac{1744}{615},
\frac{4040}{369}\right),  $$
 \begin{equation}\label{HJ1}
-H^{-1}J_0=\left( \frac{-7}{4}, \frac{-7}{4}, \frac{-7}{4},
\frac{-7}{4}, \frac{143}{10}\right).
 \end{equation}
$$-\frac{1}{2}J_0 H^{-1} J_0^T =\frac{126118}{1845} \approx 68.35664.$$
Note that the last coordinate
of $-H^{-1}J_0$ (in (\ref{HJ1})) is positive and hence we have
equality in Lemma~\ref{lem:mai}.  Thus,
\[
{\cal L}_{S}(\mu_x)
 =
 \mu^T\ln\mu + xh(v)
 +   x^2\frac{126118}{1845}
+ O(x^3)
\]

The largest increase
in likelihood is attained on $T_1$.
 Thus $T_1$ is the tree with highest
likelihood for sufficiently small $x$.

\subsection{Proof of Theorem \ref{thm:MLEbad} in CFN: Arbitrary $C$}

The Hessian is the same for all  four-leaf trees.  In this case we state the inverse of the
Hessian which is simpler to state than the Hessian itself.
We have
$$
H^{-1}=\begin{pmatrix}
\frac{16C^4-1}{32C^2} & 0 & 0 & 0 & \frac{(4C^2-1)^2}{128C^3} \vspace{0.1cm}\\
0 & \frac{16C^4-1}{32C^2} & 0 & 0 & \frac{(4C^2-1)^2}{128C^3} \vspace{0.1cm}\\
0 & 0 & \frac{16C^4-1}{32C^2} & 0 & \frac{(4C^2-1)^2}{128C^3} \vspace{0.1cm}\\
0 & 0 & 0 & \frac{16C^4-1}{32C^2} & \frac{(4C^2-1)^2}{128C^3} \vspace{0.1cm}\\
\frac{(4C^2-1)^2}{128C^3} & \frac{(4C^2-1)^2}{128C^3} &
\frac{(4C^2-1)^2}{128C^3} & \frac{(4C^2-1)^2}{128C^3} &
\frac{(16 C^4 - 24C^2 - 3)(4C^2-1)^2}
{256C^4(1+4C^2)^2}
\end{pmatrix}
$$
The first $4$ coordinates of the Jacobians are equal and have
the same value for all trees:
$$
\frac{16C(64C^6-16C^4+12C^2+1)}
{(16C^4-1)(16C^4+24C^2+1)}.
$$
Thus for each tree we will only list the last coordinate of the
Jacobian $J_0$ which we denote as $J_0[5]$.

Let
\begin{eqnarray*}
\beta&=&(4C^2-1)^2(16C^4+24C^2+1),\\
\gamma&=&\beta\cdot\frac{(1+4C^2)^2}{16}
\end{eqnarray*}
Note that $\beta>0$ and $\gamma>0$ for $C\in (0,1/2)$.

For $T_1$ we have
$$
J_0[5]=\frac{1}{\beta}\cdot 128C^2(16C^6-24C^4+17C^2+1),
$$
$$
\Delta_1:=\frac{1}{2}J_0H^{-1}J_0=\frac{1}{\gamma}\cdot(
 -512 C^{12}   - 2048 C^{10}   + 3520 C^8  - 1856 C^6  + 390 C^4  + 88 C^2  +
 3),
$$
and the last coordinate of $-H^{-1}J_0$ is
$$
L:=\frac{48C^6-40C^4+15C^2+2}{2C^2(1+4C^2)^2}.
$$
It is easily checked that $L$ is positive for $C\in(0,1/2)$ and
hence we have equality in Lemma~\ref{lem:mai}.

For $T_2$ we have
$$
J_0[5]=\frac{1}{\beta}\cdot 128C^4(16C^4-40C^2-7),
$$
$$
\Delta_2:=\frac{1}{2}J_0H^{-1}J_0=\frac{1}{\gamma}\cdot (
 -512 C^{12}   - 5120 C^{10}   + 960 C^8  + 832 C^6  + 198 C^4  + 28 C^2  +
 1).
$$

For $T_3$ we have
$$
J_0[5]=\frac{1}{\beta}\cdot 256C^4(16C^4-8C^2-3),
$$
$$
\Delta_3:=\frac{1}{2}J_0H^{-1}J_0= \frac{1}{\gamma}\cdot (
 2048 C^{12}   - 2048 C^{10}   - 512 C^8  + 72 C^4  + 24 C^2  +
 1).
$$

It remains to show that $\Delta_1>\Delta_2$ and
$\Delta_1>\Delta_3$. We know that for $C=1/4$ the inequalities
hold. Thus we only need to check that $\Delta_1-\Delta_2$
and $\Delta_1-\Delta_3$ do not have roots for $C\in (0,1/2)$.
This is easily done using Sturm sequences, which is
a standard approach for counting the number of roots of
a polynomial in an interval.

\subsection{Proof of Theorem \ref{thm:MLEbad} in JC, K2, and K3}

Our technique requires little additional  work to extend the
result to JC, K2, and K3 models. Let JC-likelihood of tree $T$ on
distribution $\mu$ be the maximal likelihood of $\mu_{T,w}$ over
all labelings of $w$, in the JC model. Similarly we define
K2-likelihood and K3-likelihood. Note that K3-likelihood of a tree
is greater or equal to its K2-likelihood which is greater or equal
to its JC-likelihood.  In the following we will consider a mixture distribution
generated from the JC model,  and look at the likelihood (under a non-mixture)
for JC, K2 and K3 models.

For the K3 model, the transition matrices are of the form
$$
P_{\rm K3}(\alpha,\beta,\gamma)=\left(
\begin{array}{cccc}
1-\alpha-\beta-\gamma & \alpha & \beta & \gamma \\
\alpha & 1-\alpha-\beta-\gamma & \gamma & \beta \\
\beta & \gamma & 1-\alpha-\beta-\gamma & \alpha \\
\gamma & \beta & \alpha & 1-\alpha-\beta-\gamma
\end{array}
\right), $$ with $\alpha\geq \beta\geq \gamma>0$, $\alpha+\beta<1/2$, and $\gamma>(\alpha+\gamma)(\beta+\gamma)$.
The K2 model is the case $\beta=\gamma$, the JC model is the case $\alpha=\beta=\gamma$.

\begin{theorem}\label{MLEbadOther} Let
$\vec{P_1}=(1/8+x,1/8-x,1/8-x,1/8+x,x^2)$ and $\vec{P_2}=(1/8-x,1/8+x,1/8+x,1/8-x,x^2)$.
Let $\mu_x$ denote the following mixture distribution on $T_3$ generated
from the JC model:  \[\mu_x=\left(\mu_{T_3,\vec{P_1}}+\mu_{T_3,\vec{P_2}}\right)/2.\]
There exists $x_0>0$ such that for all $x\in(0,x_0)$ the
JC-likelihood of $T_1$ on $\mu_x$ is higher than the
K3-likelihood of $T_2$ and $T_3$ on $\mu_x$.
\end{theorem}

Note, Part \ref{thm:JCbad} of Theorem~\ref{thm:MLEbad} for the JC, K2, and K3 models
is immediately implied by Theorem~\ref{MLEbadOther}.

First we argue that in the JC model $T_1$ is the most likely tree.
As in the case for the CFN model, because of symmetry, we
have the same Hessian for all trees.
$$H=\begin{pmatrix}
\frac{-2373504}{112255}  & \frac{-915872}{336765}  & \frac{-915872}{336765}  & \frac{-915872}{336765}  & \frac{-587856}{112255} \vspace{0.1cm}\\
\frac{-915872}{336765}  & \frac{-2373504}{112255}  & \frac{-915872}{336765}  & \frac{-915872}{336765}  & \frac{-587856}{112255} \vspace{0.1cm}\\
\frac{-915872}{336765}  & \frac{-915872}{336765}  & \frac{-2373504}{112255}  & \frac{-915872}{336765}  & \frac{-587856}{112255} \vspace{0.1cm}\\
\frac{-915872}{336765}  & \frac{-915872}{336765}  & \frac{-915872}{336765}  & \frac{-2373504}{112255}  & \frac{-587856}{112255} \vspace{0.1cm}\\
\frac{-587856}{112255}  & \frac{-587856}{112255}  & \frac{-587856}{112255}  & \frac{-587856}{112255}  & \frac{-1130124}{112255}
\end{pmatrix},$$
Again the Jacobians differ only in the last coordinate.

For $T_3$:
\begin{equation}\label{J3}
J_0=\left( \frac{4199248}{112255},
\frac{4199248}{112255}, \frac{4199248}{112255},
\frac{4199248}{112255}, \frac{-7085428}{112255}\right),\end{equation}
Then,
$$-\frac{1}{2}J_0 H^{-1} J_0^T =\frac{174062924259638}{237159005655}
\approx 733.9503.$$

For $T_2$:
\begin{equation}\label{J2}
J_0=\left( \frac{4199248}{112255}, \frac{4199248}{112255},
\frac{4199248}{112255}, \frac{4199248}{112255},
\frac{-8069818}{112255}\right),
\end{equation}
Then,
$$-\frac{1}{2}J_0 H^{-1}
J_0^T =\frac{410113105846051}{474318011310} \approx 864.6374.$$

For $T_1$:
\begin{equation}\label{J1}
J_0=\left( \frac{4199248}{112255}, \frac{4199248}{112255},
\frac{4199248}{112255}, \frac{4199248}{112255},
\frac{22878022}{112255}\right),\end{equation}
 \begin{equation}\label{HJ11}-H^{-1}J_0=\left(
\frac{-10499073}{2816908}, \frac{-10499073}{2816908},
\frac{-10499073}{2816908}, \frac{-10499073}{2816908},
\frac{118305233}{4225362}\right).
\end{equation}
Note, again the last coordinate is positive as required for the application of
Lemma~\ref{lem:mai}.
Finally,
\begin{equation}
\label{T1:JC}
-\frac{1}{2}J_0 H^{-1}
J_0^T =\frac{1221030227753251}{474318011310} \approx 2574.286.
\end{equation}

Now we bound the K3-likelihood of
$T_2$ and $T_3$. The Hessian matrix is now $15\times 15$. It
is the same for all the $4$-leaf trees and has a lot of symmetry.
There are only $8$ different entries in $H$. For distinct $i,j\in [4]$ we have
\begin{eqnarray*}
\frac{\partial^2}{\partial p_i\partial
p_i}f(\mu_0)&=&-538996/112255,
\\
\frac{\partial^2}{\partial p_i\partial
p_j}f(\mu_0)&=&-605684/1010295, \\
\frac{\partial^2}{\partial p_i\partial
p_5}f(\mu_0)&=&-132304/112255,
\end{eqnarray*}
\begin{eqnarray*}
\frac{\partial^2}{\partial p_i\partial
r_i}f(\mu_0)&=&-126086/112255, \\
\frac{\partial^2}{\partial p_i\partial
r_j}f(\mu_0)&=&-51698/336765, \\
\frac{\partial^2}{\partial p_i\partial
r_5}f(\mu_0)&=&-2448/8635,
\end{eqnarray*}
\begin{eqnarray*}
\frac{\partial^2}{\partial r_i\partial
r_i}f(\mu_0)&=&-268544/112255,
\\
\frac{\partial^2}{\partial r_i\partial
r_5}f(\mu_0)&=&-54082/112255.
\end{eqnarray*}
For $T_3$, its Jacobian $J_0$ is a vector of 15 coordinates.  It turns out that
$3J_0$ is the
concatenation of 3 copies of the Jacobian for the JC model which is stated in
\eqref{J3}.
Finally, we obtain
\begin{equation}
\label{T3:K3}
-\frac{1}{2}J_0 H^{-1} J_0^T =\frac{174062924259638}{237159005655} \approx 733.9503.
\end{equation}

For $T_2$ we again obtain that for its Jacobian $J_0$, $3J_0$ is the
concatenation of $3$ copies of \eqref{J2}.
Then,
\begin{equation}
\label{T2:K3}
-\frac{1}{2}J_0 H^{-1} J_0^T =\frac{410113105846051}{474318011310} \approx 864.6374.
\end{equation}

Finally, for $T_1$, for its Jacobian $J_0$, $3J_0$ is
the concatenation of $3$ copies of (\ref{J1}) and
$-H^{-1}J_0$ is the concatenation of $3$ copies of \eqref{HJ11}.
Then,
$$-\frac{1}{2}J_0 H^{-1} J_0^T =\frac{1221030227753251}{474318011310} \approx 2574.286.$$

Note the quantities $-\frac{1}{2}J_0 H^{-1} J_0^T $
are the same in the K3 model are the same as the corresponding quantities
in the JC model.  It appears that even though the optimization is over the
K3 parameters, the optimum assignment is a valid setting for the JC model.

Observe that
$-\frac{1}{2}J_0 H^{-1} J_0^T$ for $T_1$ in the JC model (see \eqref{T1:JC}) is larger than
for $T_2$ and $T_3$ in the $K3$ model (see \eqref{T2:K3} and \eqref{T3:K3}).
Applying Lemma \ref{lem:mai},
this completes the proof of Theorem \ref{MLEbadOther}.

\section{MCMC Results}
\label{sec:mcmc}

The following section has a distinct perspective from the earlier sections.
Here we are generating $N$ samples from the distribution and looking at
the complexity of reconstructing the phylogeny.  The earlier sections
 analyzed properties of the generating distribution, as opposed to
samples from the distribution.  In addition, instead of finding the maximum
likelihood tree, we are looking at sampling from the posterior distribution
over trees.  To do this, we consider a Markov chain whose stationary
distribution is the posterior distribution, and analyze the chain's mixing time.

For a set of data $\Dvec=(D_1,\dots,D_N)$ where $D_i\in\{0,1\}^n$,
its likelihood on tree $T$ with transition matrices $\vec{P}$ is
\begin{eqnarray*}
\mu_{T,\vec{P}}(\Dvec) &=&
\ProbCond{\Dvec}{T,\vec{P}}
\\  & = &  \prod_i
 \ProbCond{\Dvec_i}{T,\vec{P}}
 \\ &=& \prod_i
 \mu_{T,\vec{P}}(\Dvec_i)
 \\ &=& \exp(
 \sum_i \ln(\mu_{T,\vec{P}}(\Dvec_i))
\end{eqnarray*}

Let $\Phi(T,\vec{P})$ denote a prior density on the space of
trees where
\[
\sum_{T}\int_{\vec{P}} \Phi(T,\vec{P})d\vec{P} = 1.
\]
Our results extend to priors that are lower bounded by
some $\epsilon$ as in Mossel and Vigoda \cite{MV:arxiv}.
In particular, for all $T,\vec{P}$, we require $\Phi(T,\vec{P})\geq \eps$.
We refer to these priors as $\eps$-regular priors.

Applying Bayes law we get the posterior distribution:
\begin{eqnarray*}
\ProbCond{T,\vec{P}}{\Dvector}
 &=&
\frac{ \ProbCond{\Dvec}{T,\vec{P}}\Phi(T,\vec{P}) }
{\Prob{\Dvector} }
\\ & =&
\frac{ \ProbCond{\Dvec}{T,\vec{P}}\Phi(T,\vec{P}) }
{
 \sum_{T'} \int_{\vec{P}'}
  \ProbCond{\Dvec}{T',\vec{P}'}\Phi(T',\vec{P}')d\vec{P}'
  }
\end{eqnarray*}

Note that for uniform priors the posterior probability of a tree
given $\Dvec$ is proportional to $\ProbCond{\Dvec}{T}$.

Each tree $T$ then has a posterior weight
\[  w(T) = \int_{\vec{P}}
   \ProbCond{\Dvec}{T,\vec{P}}\Phi(T,\vec{P})d\vec{P}.
\]

\begin{figure}[htb]
\begin{center}
\includegraphics[type=eps,ext=.eps,read=.eps,height=2.5in]{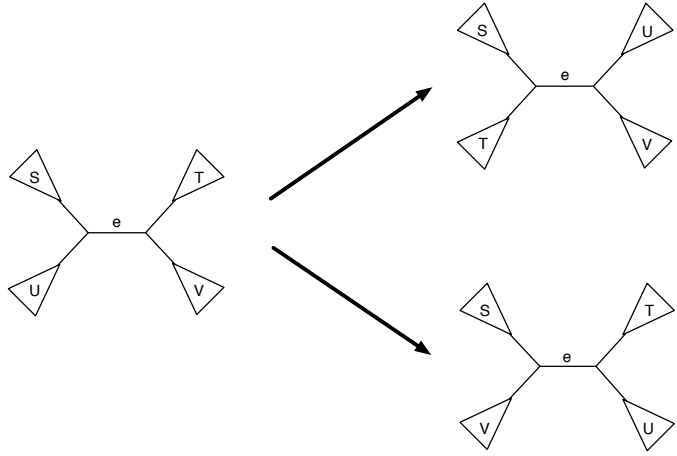}
\caption{NNI transitions.}
\label{fig:NNI}
\end{center}
\end{figure}

We look at Markov chains on the space of trees where
the stationary distribution of a tree is its posterior probability.
We consider Markov chains using nearest-neighbor interchanges (NNI).
The transitions modify the topology in the following manner which
is illustrated in Figure \ref{fig:NNI}.

Let $S_t$ denote the tree at time $t$.  The transition $S_t\rightarrow S_{t+1}$ of
the Markov chain is defined as follows.
Choose a random internal edge $e=(u,v)$ in $S$.
Internal vertices have degree three, thus let
$a,b$ denote the other neighbors of $u$ and
$y,z$ denote the other neighbors of $v$.
There are three possible assignments for these 4 subtrees to the edge $e$
(namely, we need to define a pairing between $a,b,y,z$).  Choose
one of these assignments at random, denote the new tree as $S'$.
We then set $S_{t+1}=S'$ with
 probability
\begin{equation}
\label{metrop}  \min\{1,w(S')/w(S_t)\}.
\}
\end{equation}
With the remaining probability, we set $S_{t+1}=S_t$.

The acceptance probability in \eqref{metrop} is known as the Metropolis filter
and implies that the unique stationary distribution $\pi$ of the Markov chain statisfies,
for all trees $T$:
\[  \pi(T) = \frac{w(T)}{\sum_{T'}w(T')}.
\]

We refer readers to Felsenstein \cite{F:book} and Mossel and Vigoda \cite{MV:arxiv}
for a more detailed
introduction to this Markov chain.   We are interested in the mixing time $T_{\rm{mix}}$,
defined as the number of steps until the chain is within variation distance $\leq 1/4$
of the stationary distribution.  The constant $1/4$ is somewhat arbitrary, and
can be reduced to any $\delta$ with $T_{\rm{mix}}\log(1/\delta)$ steps.

For the MCMC result we consider trees on 5 taxa.
Thus trees in this section will have leaves numbered $1,2,3,4,5$ and
internal vertices numbered $6,7,8$.
Let $S_3$ denote
the tree $(((12),5),(34))$.
Thus, $S_3$ has edges
$e_1=\{1,6\}$, $e_2=\{2,6\}$, $e_3=\{3,8\}$, $e_4=\{4,8\}$,
$e_5=\{5,7\}$, $e_6=\{6,7\}$, $e_7=\{7,8\}$.  We will list the transition probabilities for
the edges of $S_3$ in this order.
For the CFN model, we consider the following vector of transition probabilities.
Let
\begin{eqnarray*}
\vec{P_1} =&(1/4+x,1/4-x,1/4-x,1/4+x,1/4,c,c),& \mbox{ and } \\
\vec{P_2} =&(1/4-x,1/4+x,1/4+x,1/4-x,1/4,c,c),
\end{eqnarray*}
where
$$c=
\frac{1}{2}\left(1 - \sqrt{\frac{1-16x^2}{1+16x^2} }\right).
$$

For the JC model, let
\begin{eqnarray*}
\vec{P_1}=&(1/8+x,1/8-x,1/8-x,1/8+x,1/8,c',c'),& \mbox{ and } \\
\vec{P_2}=&(1/8-x,1/8+x,1/8+x,1/8-x,1/8,c',c'),
\end{eqnarray*}
where
$$c'=16x^2.
$$

Let $\mu_1=\mu_{S_1,\vec{P_1}}$ and $\mu_2=\mu_{S_1,\vec{P_2}}$.
We are interested in the mixture distribution:
\[  \mu = \frac{1}{2}\left(\mu_1+\mu_2\right)
\]
Let $S_2$ denote the tree $(((14),5),(23))$.

The key lemma for our Markov chain result states that under $\mu$,
the likelihood has local maximum, with respect to NNI connectivity,
on $S_1$ and $S_2$.
\begin{lemma}
\label{lem:MCMC-MLE}
For the CFN and JC models, there exists $x_0>0$ such that for
all $x\in(0,x_0)$ then for all trees $S$ that are one NNI transition from
$S_1$ or $S_2$, we have
\[
{\cal L}_S(\mu) < {\cal L}_{S_1}(\mu),
\ \ \
{\cal L}_S(\mu) < {\cal L}_{S_2}(\mu)
\]
\end{lemma}

This then implies the following corollary.
\begin{theorem}
\label{thm:MCMC-main}
There exist a constant $C > 0$
such that for all $\eps>0$ the following holds.
Consider a data set with $N$ characters, i.e., $\vecD=(D_1,\dots,D_N)$, chosen
independently from the distribution $\mu$.
Consider the Markov chains on tree topologies
defined by nearest-neighbor interchanges (NNI).
Then with probability $1 - \exp(-C  N)$ over the data generated,
the mixing time of the Markov chains, with priors which are $\eps$-regular,
satisfies
\[ T_{mix} \geq  \eps\exp(C N).
\]
\end{theorem}

The novel aspect of this section is Lemma~\ref{lem:MCMC-MLE}.
The proof of Theorem~\ref{thm:MCMC-main} using Lemma~\ref{lem:MCMC-MLE}
is straightforward.

\begin{proof}[Proof of Lemma \ref{lem:MCMC-MLE}]
The proof follows the same lines as
the proof of Theorems \ref{thm:MLEbad} and \ref{MLEbadOther}.
Thus our main task is to compute the Hessian and Jacobians, for
which we utilize Maple.

We begin with the CFN model.

The Hessian is the same for all $15$ trees on 5-leaves:
$$H=
\begin{pmatrix}
\frac{-3880}{1281}  & \frac{-114}{427}  & \frac{-114}{427}  &
\frac{-114}{427}  & \frac{-114}{427}  & \frac{-1241}{1281}  &
\frac{-190}{427} \vspace{0.1cm}\\
\frac{-114}{427}  & \frac{-3880}{1281}  & \frac{-114}{427}  &
\frac{-114}{427}  & \frac{-114}{427}  & \frac{-1241}{1281}  &
\frac{-190}{427} \vspace{0.1cm}\\
\frac{-114}{427}  & \frac{-114}{427}  & \frac{-3880}{1281}  &
\frac{-114}{427}  & \frac{-114}{427}  & \frac{-190}{427}  &
\frac{-1241}{1281} \vspace{0.1cm}\\
\frac{-114}{427}  & \frac{-114}{427}  & \frac{-114}{427}  &
\frac{-3880}{1281}  & \frac{-114}{427}  & \frac{-190}{427}  &
\frac{-1241}{1281} \vspace{0.1cm}\\
\frac{-114}{427}  & \frac{-114}{427}  & \frac{-114}{427}  &
\frac{-114}{427}  & \frac{-3880}{1281}  & \frac{-190}{427}  &
\frac{-190}{427} \vspace{0.1cm}\\
\frac{-1241}{1281}  & \frac{-1241}{1281}  & \frac{-190}{427}  &
\frac{-190}{427}  & \frac{-190}{427}  & \frac{-6205}{3843}  &
\frac{-950}{1281} \vspace{0.1cm}\\
\frac{-190}{427}  & \frac{-190}{427}  & \frac{-1241}{1281}  &
\frac{-1241}{1281}  & \frac{-190}{427}  & \frac{-950}{1281}  &
\frac{-6205}{3843}
\end{pmatrix},$$

We begin with the two trees of interest: \\

\treee{2}{3}{5}{4}{1} \ \ \ \ \ and\ \ \ \ \treee{1}{2}{5}{4}{3}

Their Jacobian is
$$ J_0=\left(
\frac{17056}{1281},
\frac{17056}{1281},
\frac{17056}{1281},
\frac{17056}{1281},
\frac{2432}{427},
\frac{57952}{3843},
\frac{57952}{3843}\right),$$
Thus,
$$-H^{-1}J_0=\left(
2,
2,
2,
2,
0,
4,
4\right).$$
Note the last two coordinates are positive, hence
we get equality in the conclusion of Lemma~\ref{lem:mai}.
Finally,
$$-\frac{1}{2}J_0 H^{-1} J_0^T =\frac{436480}{3843} \approx 113.5779.$$

We now consider those trees connected to $S_1$ and $S_2$ by one
NNI transition.  Since there are 2 internal edges each tree has 4 NNI
neighbors.

The neighbors of $S_1$ are

\treee{3}{4}{2}{5}{1}
\treee{3}{4}{1}{5}{2}
\treee{1}{2}{3}{5}{4}
\treee{1}{2}{4}{5}{3}

The neighbors of $S_2$ are

\treee{2}{3}{1}{5}{4}
\treee{2}{3}{4}{5}{1}
\treee{1}{4}{2}{5}{3}
\treee{1}{4}{3}{5}{2}

The Jacobian for all 8 of these trees is
$$J_0=\left(
\frac{17056}{1281},
\frac{17056}{1281},
\frac{17056}{1281},
\frac{2432}{427},
\frac{17056}{1281},
\frac{57952}{3843},
\frac{3728}{427}\right),$$
Finally,
$$-\frac{1}{2}J_0 H^{-1} J_0^T =\frac{2242633984}{20840589} \approx
107.6090,$$

Note the quantities $-\frac{1}{2}J_0 H^{-1} J_0^T$ are larger for the
two trees $S_1$ and $S_2$.  This completes the proof for the CFN model.

We now consider the JC model.
Again the Hessian is the same for all the trees:
$$H=\begin{pmatrix}
\frac{-512325018}{20541185}  & \frac{-36668964}{20541185}  &
\frac{-36668964}{20541185}  & \frac{-36668964}{20541185}  &
\frac{-36668964}{20541185}  & \frac{-328636353}{41082370}  &
\frac{-28145979}{8216474} \vspace{0.1cm}\\
\frac{-36668964}{20541185}  & \frac{-512325018}{20541185}  &
\frac{-36668964}{20541185}  & \frac{-36668964}{20541185}  &
\frac{-36668964}{20541185}  & \frac{-328636353}{41082370}  &
\frac{-28145979}{8216474} \vspace{0.1cm}\\
\frac{-36668964}{20541185}  & \frac{-36668964}{20541185}  &
\frac{-512325018}{20541185}  & \frac{-36668964}{20541185}  &
\frac{-36668964}{20541185}  & \frac{-28145979}{8216474}  &
\frac{-328636353}{41082370} \vspace{0.1cm}\\
\frac{-36668964}{20541185}  & \frac{-36668964}{20541185}  &
\frac{-36668964}{20541185}  & \frac{-512325018}{20541185}  &
\frac{-36668964}{20541185}  & \frac{-28145979}{8216474}  &
\frac{-328636353}{41082370} \vspace{0.1cm}\\
\frac{-36668964}{20541185}  & \frac{-36668964}{20541185}  &
\frac{-36668964}{20541185}  & \frac{-36668964}{20541185}  &
\frac{-512325018}{20541185}  & \frac{-28145979}{8216474}  &
\frac{-28145979}{8216474} \vspace{0.1cm}\\
\frac{-328636353}{41082370}  & \frac{-328636353}{41082370}  &
\frac{-28145979}{8216474}  & \frac{-28145979}{8216474}  &
\frac{-28145979}{8216474}  & \frac{-1273864167}{82164740}  &
\frac{-134747901}{20541185} \vspace{0.1cm}\\
\frac{-28145979}{8216474}  & \frac{-28145979}{8216474}  &
\frac{-328636353}{41082370}  & \frac{-328636353}{41082370}  &
\frac{-28145979}{8216474}  & \frac{-134747901}{20541185}  &
\frac{-1273864167}{82164740}
\end{pmatrix}$$

Beginning with tree $S_1$

\treee{1}{2}{5}{4}{3}

We have:
$$J_0=\left(
\frac{4342624176}{20541185},
\frac{4342624176}{20541185},
\frac{4342624176}{20541185},
\frac{4342624176}{20541185},
\frac{1733695536}{20541185},
\frac{5655197244}{20541185},
\frac{5655197244}{20541185}\right),$$
The last two coordinates of $-H^{-1}J_0$ are
$$
\frac{5114490004637540016}{593018923302763639},
\frac{5114490004637540016}{593018923302763639}.$$
Since they are positive we get equality in the conclusion of Lemma \ref{lem:mai}.
Finally,
$$-\frac{1}{2}J_0 H^{-1} J_0^T
=\frac{48101472911555370428804991552}{12181311412062878919972215}
\approx 3948.793$$

For the neighbors of $S_1$:

\treee{3}{4}{1}{5}{2}
\treee{3}{4}{2}{5}{1}
\treee{1}{2}{3}{5}{4}
\treee{1}{2}{4}{5}{3}

We have:
$$J_0=\left(
\frac{4342624176}{20541185},
\frac{4342624176}{20541185},
\frac{4342624176}{20541185},
\frac{1733695536}{20541185},
\frac{4342624176}{20541185},
\frac{5655197244}{20541185},
\frac{2955839412}{20541185}\right).$$
Hence,
$$-\frac{1}{2}J_0 H_1^{-1} J_0^T
=\frac{56725804836101083569837263821270061643565080096}{15568481282665727860752794372508821870798435}
\approx 3643.631$$

Considering $S_2$:

\treee{2}{3}{5}{4}{1}

$$J_0=\left(
\frac{4342624176}{20541185},
\frac{4342624176}{20541185},
\frac{4342624176}{20541185},
\frac{4342624176}{20541185},
\frac{1733695536}{20541185},
\frac{1074039432}{4108237},
\frac{1074039432}{4108237}\right),$$
The last two coordinates of $-H^{-1}J_0$ are
$$\frac{13458396942990580792}{1779056769908290917},
\frac{13458396942990580792}{1779056769908290917},$$
which are positive.  Finally,
$$-\frac{1}{2}J_0 H^{-1} J_0^T
=\frac{45365294744197291555715368032}{12181311412062878919972215}
\approx 3724.172$$

Considering the neighbors of $S_2$:

\treee{1}{4}{2}{5}{3}
\treee{1}{4}{3}{5}{2}
\treee{2}{3}{1}{5}{4}
\treee{2}{3}{4}{5}{1}

We have:
$$J_0=\left(
\frac{4342624176}{20541185},
\frac{4342624176}{20541185},
\frac{4342624176}{20541185},
\frac{1733695536}{20541185},
\frac{4342624176}{20541185},
\frac{1074039432}{4108237},
\frac{2955839412}{20541185}\right)$$
Hence,
$$-\frac{1}{2}J_0 H^{-1} J_0^T
=\frac{7756149367472421142972629871553505755962112808}{2224068754666532551536113481786974552971205}
\approx 3487.369$$

\end{proof}

\begin{remark}
In fact $S_1$ and $S_2$ have larger likelihood than any of the 13 other
5-leaf trees.  However, analyzing the likelihood for the 5 trees not considered
in the proof of Lemma \ref{lem:MCMC-MLE} requires more
technical work since $-\frac{1}{2}J_0 H^{-1} J_0^T$ is maximized at
invalid weights for these 5 trees.
\end{remark}

We now show how the main theorem of this section easily follows from the above lemma.
The proof follows the same basic line of argument as in Mossel and Vigoda \cite{MV}, we point
out the few minor differences.

\begin{proof}[Proof of Theorem \ref{thm:MCMC-main}]

For a set of characters $\vec{D}=(D_1,\dots,D_N)$, define the maximum log-likelihood of
tree $T$ as
\[
{\cal L}_{T}(\vec{D})  = \max_{\vec{P}\in{\mathcal{M}}^E} {\cal L}_{T,\vec{P}}(\vec{D}),
\]
where
\[
 {\cal L}_{T,\vec{P}}(\vec{D}) = \sum_{i} \ln\left(\ProbCond{D_i}{T,\vec{P}}\right).
 \]

Consider $\vec{D}=(D_1,\dots,D_N)$ where each $D_i$ is independently sampled from the mixture
distribution $\mu$.
Let $S^*$ be $S_1$ or $S_2$, and let $S$ be a tree that is one NNI transition from $S^*$.
Our main task is to show that ${\cal L}_{S^*}(\vec{D}) > {\cal L}_{S}(\vec{D})$.

Let $\vec{P^*}$ denote the assignment which attains the maximum for ${\cal L}_{S^*}(\mu)$,
and let
\[   \alpha=\min_{\sigma\in\Omega^5} \ProbCond{\sigma}{S^*,\vec{P^*}}.  \]

   For $\sigma\in\Omega^5$, let $D(\sigma)=|\{i:D_i=\sigma\}|$.
By Chernoff's inequality (e.g., \cite[Remark 2.5]{JLR}), and a union bound over $\sigma\in\Omega^5$,
we have for all $\delta>0$,
\[ \Prob{ \mbox{ for all } \sigma\in\Omega^5, |D(\sigma)-\mu(\sigma)N| \leq \delta N }
     \geq 1 - 2\cdot 4^5\exp(-2\delta^2N) = 1 - \exp(-\Omega(N)).
 \]
 Assuming the above holds, we then have
 \[
 {\cal L}_{S}(\vec{D})  \leq  N(1-\delta){\cal L}_{S}(\mu).
 \]
 And,
 \[
{\cal L}_{S^*}(\vec{D}) \geq  {\cal L}_{S^*,\vec{P^*}}(\vec{D}) \geq  N(1-\delta){\cal L}_{S^*}(\mu) + 4^5\delta N\log(\alpha)
 \]
  Let $\gamma := {\cal L}_{S^*,\vec{P^*}}(\mu) - {\cal L}_{S,\vec{P}}(\mu).$
  Note, Lemma \ref{lem:MCMC-MLE} states that $\gamma>0$.

  Set $\delta  := \frac{\min\{1,\gamma/10\}}{4^5\log(1/\alpha)}$.  Note, $4^{-5}>\delta>0$.
Hence,
\[
{\cal L}_{S^*}(\vec{D}) - {\cal L}_{S}(\vec{D}) \geq
    N(1-\delta)\gamma - N\gamma/10
      > N\gamma/5 = \Omega(N).
 \]

 This then implies that $w(S)/w(S^*) \leq \exp(\Omega(N))$ with probability $\geq 1-\Omega(N)$
 by the same argument as the proof of
 Lemma 21 in Mossel and Vigoda \cite{MV}.  Then, the theorem follows from a conductance
 argument as in Lemma 22 in \cite{MV}.

 \end{proof}

\section{Acknowledgements}

We thank Elchanan Mossel for useful discussions, and John Rhodes for helpful comments
on an early version of this paper.

\end{document}